\documentclass[fleqn,usenatbib]{mnras}

\usepackage{newtxtext,newtxmath}
\usepackage[title]{appendix}
\usepackage[dvipsnames]{xcolor}

\usepackage[T1]{fontenc}

\usepackage{graphicx}	
\usepackage{amsmath}	
\usepackage{comment}
\usepackage{soul}
\usepackage{bm}
\usepackage{tabularx}
\usepackage[parfill]{parskip}
\usepackage{secdot}
\usepackage{booktabs}

\title[Unsupervised learning of IllustrisTNG populations]{Applying unsupervised learning to resolve evolutionary histories and explore the galaxy-halo connection in IllustrisTNG}

    \author[T.S. Fraser et al.]{
T.S. Fraser,$^{1}$\thanks{E-mail: tsfraser@uwaterloo.ca}
R. Tojeiro,$^{2}$
H. G. Chittenden$^{2}$
\\
$^{1}$Waterloo Centre for Astrophysics, Department of Physics \& Astronomy, University of Waterloo, 200 University Ave. W., Waterloo, Ontario, Canada N2L 3G1\\
$^{2}$School of Physics \& Astronomy, University of St Andrews, North Haugh, St Andrews KY16 9SS, Scotland, United Kingdom
}

\date{Accepted 2022 December 17. Received Sept 6 2022; in original form Dec 23 2021}

\pubyear{2022}

\begin{document}
\label{firstpage}
\pagerange{\pageref{firstpage}--\pageref{lastpage}}
\maketitle
\begin{abstract}

We examine the effectiveness of identifying distinct evolutionary histories in IllustrisTNG-100 galaxies using unsupervised machine learning with Gaussian Mixture Models. We focus on how clustering compressed metallicity histories and star formation histories produces subpopulations of galaxies with distinct evolutionary properties (for both halo mass assembly and merger histories). By contrast, clustering with photometric colours fail to resolve such histories. We identify several populations of interest that reflect a variety of evolutionary scenarios supported by the literature. Notably, we identify a population of galaxies inhabiting the upper-red sequence, $M_{*} > 10^{10} M_{\odot}$ that has a significantly higher ex-situ merger mass fraction present at fixed masses, and a star formation history that has yet to fully quench, in contrast to an overlapping, satellite-dominated population along the red sequence, which is distinctly quiescent. Extending the clustering to study four clusters instead of three further divides quiescent galaxies, while star forming ones are mostly contained in a single cluster, demonstrating a variety of supported pathways to quenching. In addition to these populations, we identify a handful of populations from our other clusters that are readily applicable to observational surveys, including a population related to post starburst (PSB) galaxies, allowing for possible extensions of this work in an observational context, and to corroborate results within the IllustrisTNG ecosystem.
\newline

\end{abstract}




\section{Introduction}\label{sec:intro}

In concordance cosmology, galaxies form within dark matter halos, and follow hierarchical evolutionary histories \citep{cole2000hierarchical}.The so-called "galaxy-halo connection" is an umbrella term that attempts to capture the complex relationship between the evolution of galaxies and their halos \citep[e.g.][]{Wechsler_2018}. For example, the observed spacial distribution of galaxies is reasonably well modelled by populating dark-matter halos of different mass (and other properties, such as age) with different types of galaxies\citep[e.g.][]{Zehavi_2005, Guo_2014}, and the role of mergers - a prominent prediction of hierarchical structure formation - has long been considered in the growth of galaxies \citep[e.g.][]{Wake_2006, Bundy_2009, Tojeiro_2012, Moustakas_2013}, galaxy morphological transformation \citep[e.g.][]{Bertone_2009,Martin_2018}, the quenching of star formation \citep[e.g.][]{Pawlik_2018} or the onset of AGN \citep[e.g.][]{Villforth_2014} . However, the evolution of halos is not directly observable, and the evolution of galaxies is only partially observable: star formation and metallicity histories give key insights into major evolutionary events, but do not account for the entirety of it. Cosmological-hydrodynamic simulations have shown that star formation histories are extraordinarily sensitive to a variety of factors, and consequently produces a large scatter driven by these parameters, rendering evolutionary descriptions incomplete \citep[e.g.][]{Lagos_2016,diemer2017log,cochrane2018dissecting}. 

Simulations allow for non-observable halo and galaxy properties to be considered in the context of their evolution, e.g. dark matter, local environment, halo concentration, and halo mass assembly histories have been successfully modelled. As a consequence, as simulated galaxies begin to match observed galaxies on a range of properties and redshifts \citep[e.g.][]{Henriques_2015, Pillepich2017, 10.1093/mnras/stz937}, questions regarding the impact of halo evolution on the evolutionary histories of galaxies can be explored. Given the large parameter space that galaxy observables occupy, simulations can be used to explore emergent relationships between different parameters, which were not originally considered \citep[e.g.][]{Tojeiro_2017}.

Simulations like IllustrisTNG \cite{nelson2018illustristng,2017stellar,2017cluster,2017bimod,2018chem,2018mag}, EAGLE \citep{Schaye_2014} and SIMBA \citep{10.1093/mnras/stz937} and their associated bodies of literature present a remarkable opportunity, as they contain thousands of simulated galaxies, with a multitude of observable and non-observable parameters, governed by a vast array of non-linear, multi-variate relationships. This has led to work that explored the use of unsupervised learning techniques to uncover data-driven relationships that we would not be privy to a priori. One source of success was in applying dimensionality reduction, through the form of Principal Component Analysis (PCA), which has seen application to a wide variety of fields in astrophysics and in galaxy evolution. Dimensionality reduction can result in the discovery of physically driven mechanisms present in galaxies and allow us to quantify them. \citet{Lagos_2016}, through use of PCA, uncovered a “fundamental plane” of star forming galaxies, driven by a correlation between the star formation rate, the neutral gas fraction and the stellar mass. These three variables were seen to account for the majority of the variance seen in galaxies, and the authors were able to interpret the shape of this fundamental plane as a result of the self-regulation of the galaxies’ star formation rates.  Another example of the use of PCA in studying relationships between galaxy observables is \citet{cochrane2018dissecting}, which again employed EAGLE simulation data to study connections between the star formation rate, dark matter halo mass and stellar mass across redshifts. These connections were further examined by splitting of galaxies as central or satellites, and according to their stellar mass. This sort of analysis is pertinent for us given that it directly links the star formation history with halo parameters, and the PCA seems to suggest at a few physically driven mechanisms at work. The relative strength of this $M_*-M_h-SFR$ relation varies based on whether the galaxy's mass, central/satellite status, which implies that the driving variables of the star formation history, and their link to the halo, might vary based on the type of galaxy, in addition to highlighting differences in quenching regimes.

The scatter in star formation histories, for galaxies of similar mass, has also been studied using simulations. \citet{Sparre_2015}, using PCA on the Illustris simulation, argue that variability on scales above or below 500Myrs are likely related to different processes, with halo assembly histories responsible for long time-scale variations, and gas accretion for short time-scale effects. \citeauthor{diemer2017log} fitting SFHs using log-normals, found that the scatter in star formation histories could not be uniquely determined by a single parameter. \citet{cohn2018approximations}, building off of this work, compared \citet{diemer2017log}'s log-normal model to one developed with PCA. Although PCA is able to reproduce simulated SFHs, the authors point out that the underlying assumption of PCA - that scatter seen in the simulation is but a deviation from a single underlying star formation history - is unlikely to be true. Rather, \citeauthor{cohn2018approximations} states: “It is possible to group galaxies into more than two subfamilies, with each subfamily having similar integrated SFR histories” . Additional support for this diversity of groups of formation histories is seen in \citet{Carnall_2018}.  Taken together, these studies propose that we do not interpret formation histories and their scatter as deviations from a single, underlying, general star formation history, but instead consider the existence of a series of sub-families, each with different star formation histories, and scatter. This is where the case for unsupervised learning via clustering is strong: as small populations can contribute little to the overall variance, but can be distinct and of astrophysical interest.

This sensitivity problem is not unique to star formation histories. Halo assembly histories have consistently demonstrated a multivariate sensitivity to other parameters. \citet{wang2020concentrations} underscores the importance of parameterizing the mass assembly history (MAH), as even the occurrence of minor merger events along the evolutionary history of the halo can drastically alter its structure. They note even small amounts of accretion can alter the concentration, therein driving the scatter in the mass assembly history, further affecting other halo parameters. \citet{obreschkow2020characterizing} and \citet{chen2020relating} both emphasize the importance of using the full merger tree for understanding halo evolution, outlining a major challenge in linking the mass assembly histories to star formation histories, as even subtle differences in parameterization can have drastic effects.

The sensitivity of MAHs also affects the baryonic components of galaxies.  \citet{rey2019sensitivity} find that MAHs are enormously sensitive to initial conditions, and that this sensitivity is extremely influential on the star formation history. In particular, they propose that the larger scatter in star formation histories may be partially driven by the sensitivity of MAHs to parameters surrounding mergers.

\citet{chen2020relating} address the sheer diversity of non-observable parameters that can be used to describe dark matter halos and their mass assembly histories. They outline a variety of possible parameterizations of the mass assembly history of halos, and  uncover a tight correlation between the mass assembly history of the halo and its concentration. Their analysis finds that more than 80$\%$ of the variance is explained by the mass assembly history. As stated previously, this means that the scatter of star formation histories is not wholly determined by baryonic processes, as the halo assembly history also influences the scatter seen in the star formation histories, illustrating the role the galaxy-halo connection has on these sensitive parameters.

In this paper, we will use the Illustris TNG-100 simulation to explore a wide array of observable properties, including but not limited to star-formation histories, and use non-supervised machine learning (clustering) to identify sub-families (clusters) of galaxies. Critically, we will develop an analysis that allows us to take advantage of the observability of star formation and metallicity histories (see: \cite{https://doi.org/10.48550/arxiv.2202.01809}, and through clustering, allow us to directly identify sub-populations of interest in future surveys. This work does not rank observables used in clustering per se, but instead focuses on the populations of interest that may be extracted.  Given existing literature on the driving variables of dark matter halos, we then aim to link these identified subfamilies to different features of the mass assembly history of the dark matter halos, and other halo parameters. Our research questions are: (1) can non-supervised clustering using galaxy observables reveal distinct populations in terms of their dark matter evolution? (2) How do those populations change according to the observables being clustered? 

We consider two classes of observables: optical broadband colours, and compressed star-formation and chemical enrichment histories. Time-resolved star-formation and metallicity histories are often obtained from optical spectra, sometimes in conjunction with broadband photometry (e.g. \citealt{Tojeiro_2017, Carnall_2018, https://doi.org/10.48550/arxiv.2202.01809}). Evidence shows that the addition of spectra has an important impact on the fitting process, either by mitigating the effects of the chosen priors or parametrisation \citep{https://doi.org/10.48550/arxiv.2202.01809} or by significantly sharpening the posteriors \citep{Wild_2020}. We might therefore expect that clustering using star-formation and metallicity histories allows a more detailed separation of galaxy populations.

This organized as as follows: section \ref{sec:ML} will detail the conceptual framework of unsupervised learning, and the specific techniques used in this work. Section \ref{sec:Methods} will describe the data pipeline used to extract IllustrisTNG-100 data, and the subsequent steps needed for use in our clustering and describe the motivation for these steps, as well as the framework used to select the number of clusters. Section \ref{sec:Results} will describe the results of our clustering, and highlight the demographics of the sub-populations identified, and how these populations relate to the evolutionary history of these galaxies’ dark matter halos and merger histories. Section \ref{sec:Discussion} will describe the characteristics of the sub-populations previously identified and place these results in context of the literature and discuss the role of different datasets. Section \ref{sec:Conclusion} will present the main conclusions of this work, and outline areas that are promising for future work, in both the context of existing simulations and with observational data.

\section{Machine Learning Framework}\label{sec:ML}

 Unsupervised learning was favoured here over more traditional forms of machine learning (e.g. supervised) owing to the transferability of results from observables to non-observables. By clustering, cluster membership can then be cross-referenced with dark matter halo properties of the IllustrisTNG simulation, without ever using them as a clustering variable, allowing us to probe the link between observables and dark matter halo features. 
 
 Clustering algorithms are effective at identifying distinct populations within a dataset, defined here as populations that separate in a given parameter space. It is chosen over, e.g. PCA, for its ability to identify small distinct populations that may not contribute meaningfully to dataset variance, but might otherwise be of interest to us.

\subsection{Dimensionality Reduction}

The primary objective of dimensionality reduction is to represent a large, multivariate dataset in a lower dimensional space. This is primarily achieved by constructing a basis that will adequately represent a $D$-dimensional dataset in $d$-dimensions. Two common methods of achieving this are via PCA and NMF (non-negative matrix factorization), both of which are methods that decompose the data into a series of basis vectors.

The use of dimensionality reduction with our IllustrisTNG data was strongly motivated by our interest in the star formation and metallicity histories and their roles in determining different sub-populations. The distribution of stellar ages and their metallicity can be recovered from the simulation with almost arbitrarily high resolution, making the problem intractable from a clustering perspective. At the same time, \cite{https://doi.org/10.48550/arxiv.2202.01809} demonstrate the validity of using compressed star-formation and metallicity histories as the basis vectors for a non-parametric model of the spectral energy distribution of galaxies. NMF coefficients are being directly fitted to 10 million low-redshift galaxies observed by the Dark Energy Spectroscopic Survey (\cite{https://doi.org/10.48550/arxiv.2202.01809}), creating a population of real galaxies with which our results can be compared. We use NMF components to represent the star formation and metallicity histories, and smoothed on scales of 414 Myrs (3 bins), to construct our basis.

In addition, we use PCA to gauge the distinctness of the clusters we produce, by comparing the basis vectors. We refer the reader to \cite{shlens2014tutorial}'s tutorial and \cite{2014statistics}'s section on PCA for a detailed overview. For NMF, we refer the reader to \citet{2014statistics} and \cite{zhu2016nonnegative}, as this method has often seen use in representing galaxy SEDs, among other applications, and provides a technical overview. For both of these methods, we made use of the built-in implementations offerred by sklearn \cite{scikit-learn}.

\subsection{Expectation Maximization}
As described in section \ref{sec:Methods}, we ultimately settled on a Gaussian Mixture Model (GMM) as our clustering algorithm of choice. We made use of the implementation of GMM offered by sklearn \cite{scikit-learn}. Expectation maximization is the core method that governs clustering in several algorithms, including GMM. Here, we overview its underlying concept.

Expectation maximization works by making an initial estimate of the posterior probability (the probability of a data-vector $x_{i}$ being in cluster $j$ :$p(j|x_{i})$), which is then be used to find the corresponding parameters of the distribution. Using these parameters, the probability is updated (the expectation step).  These steps are repeated until it converges on a local maximum, which effectively means that each data vector, $x_i$ now has an associated cluster membership, $j$. In the case of GMM, $p(j|x_{i})$ is a Gaussian, but this takes on a more general form for other algorithms using EM, with starting point being the partial derivatives of the log likelihood of a given probability distribution (being Gaussian in this case).

See \citet{cohn2018approximations} for a caution against modelling individual SFHs as perturbations around a single mean.

\section{Methods}\label{sec:Methods}

IllustrisTNG is a suite of twenty cosmohydrodynamical simulations, of assorted volume, mass and resolution. All IllustrisTNG simulations assume the Planck-2015 $\Lambda$CDM cosmological model (i.e. $\Omega_m = 0.3089$, $\Omega_\Lambda = 0.6911$, $\Omega_b = 0.0486$, $H_0 = 67.74$ km/s/Mpc)
\cite{Illustris,Pillepich2017}.

The public datasets are divided into 100 "snapshots" in time, covering a redshift range of 0-20, with a mean separation in time of 137.6 Myr. Each halo has an associated merger tree, detailing the properties of the numerous progenitors of the halo at prior snapshots of the simulation \citep{Jiang, IllustrisTNG}. This work considers the mass assembly and merger history of central halos along the (Main Progenitor Branch -MPB) of the tree; i.e. the history specific to the most massive halo at each snapshot. 

IllustrisTNG is run on three volumes. The runs are denoted as: TNG-50,TNG-100 and TNG-300, which are approximately 50,100 and 300 Mpc across, respectively. TNG-300 has the largest sample of galaxies, at the cost of the lowest resolution,  suited for high mass objects. By contrast, TNG-50 is more suited for studying internal processes of galaxies and the structures of individual subhalos (\citealt{nelson2018illustristng}). We opted to use the TNG-100 simulation, balancing statistics with resolution.  IllustrisTNG incorporates supermassive black hole feedback, primordial magnetic fields, and an upgraded galaxy evolution model (see \citealt{nelson2018illustristng} for details). 

\subsection{Data extraction: Baryonic Properties}

We took a mass cut of galaxies of $\log M_{*}/M_{\odot} > 9.5$ at the final snapshot ($z=0$), giving a total of 12535 galaxies.

We extracted the ages and metallicity of all stellar particles bound to each subhalo in our sample, up to two effective radii. The star-formation histories were computed as the mass formed in each of 100 bins of lookback time, linearly spaced between 0 and the age of the Universe at $z=0$. The metallicity histories were computed by taking the the mass-weighted metallicity in the same bins. Metallicity is defined as the mass fraction in elements heavier than He. We complemented our data with TNG public catalogues; these are summarised in Table \ref{tab:Sup}.

\begin{table}
    \centering
    \caption{Table of supplementary catalogues used for variables either in the clustering, or in post-clustering analysis. Catalogs cited as advised by IllustrisTNG.  Catalog G is from \protect\cite{Rodriguez_Gomez_2015, Rodriguez_Gomez_2016, Rodriguez_Gomez_2017}, and catalog K is from \protect\cite{Nelson_2017} and sources therein. The photometry and colours from catalogue K use ugriz SDSS (rest-frame) bands and a separate entry of UJV (rest frame, but not used for this analysis). The dust model accounts for effects of obscuration, and is specifically model "C" from \protect\citet{Nelson_2017}, which is their resolved dust model. For $z=0$ this photometry is provided as a series of absolute magnitudes. }
    \begin{tabularx}{\columnwidth}{c|XX}
         Catalogue index & Catalogue name & Variables used  \\ \hline
         G & Stellar assembly & $f_{InSit}$,$f_{Outgal}$,$f_{M}$, $f_{Stripped}$ \\  \hline
         K & SDSS photometry, colours and mock fiber spectra & SDSS $ugri$ magnitudes \\
    \end{tabularx}
    \label{tab:Sup}
\end{table}

\subsection{Pre-processing}\label{sec:PreP}

\begin{figure*}
    \centering
    \includegraphics[width = \textwidth]{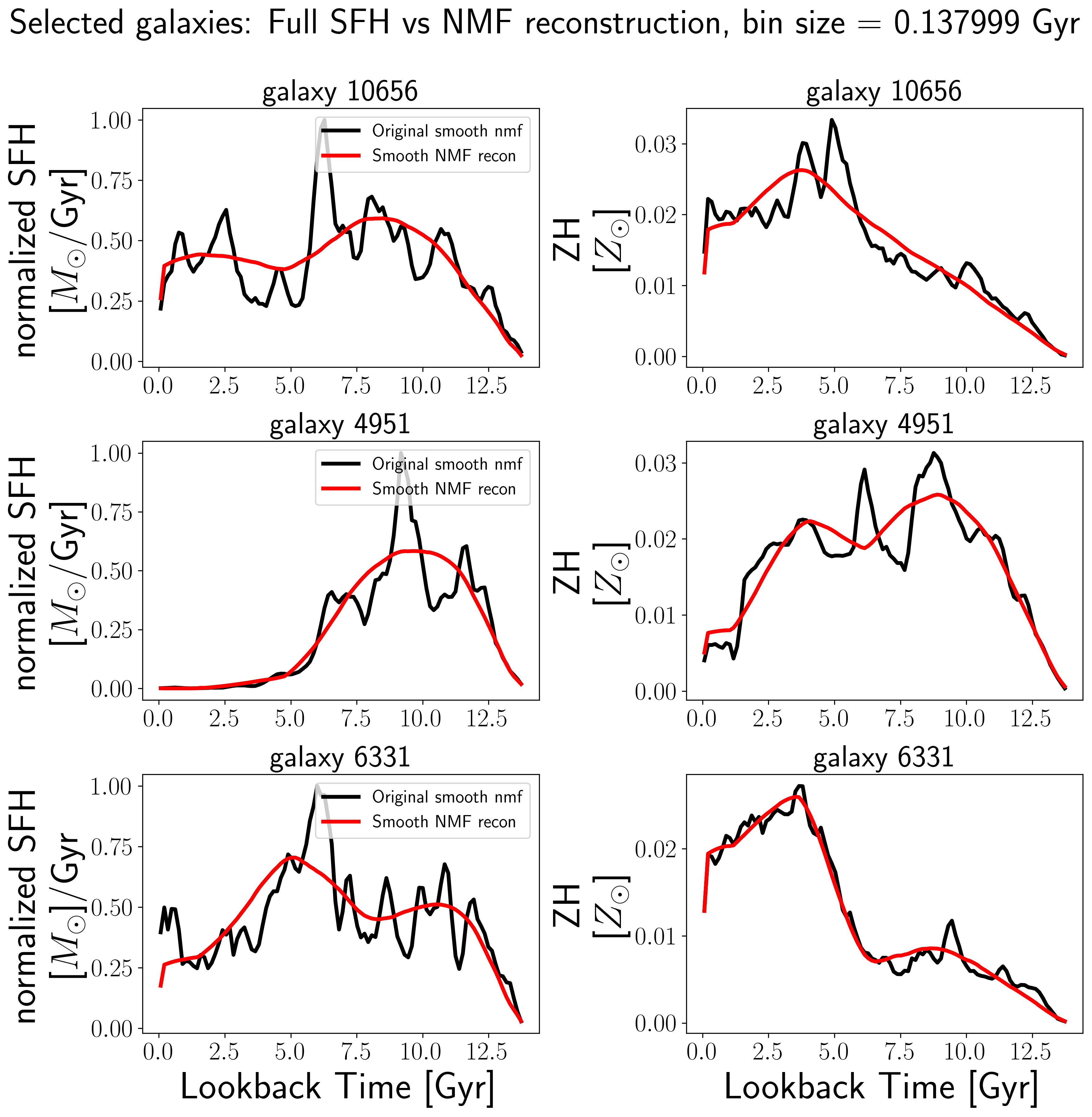}
    \caption{Plot of three reconstructed star formation histories and metallicity histories, smoothed and decomposed into four NMF components. Generally the overall behaviour of the star formation histories is captured, and the integrated stellar mass of the reconstruction is consistent with the original integrated stellar mass of Figure \ref{fig:recon_fig}.}
    \label{fig:recon_SFH}
\end{figure*}

 We compressed the star formation histories (SFH)  using NMF, in order to reduce the dimensionality of our data. We prefer to use NMF over, for example, PCA, due to the non-negativity constraints on the basis vectors. Doing so allows us to interpret basis vectors physically, as stellar mass or metallicity.

To ensure the NMF preserves the major features and relationships in the star formation histories, we normalize the star formation using a Maximum Absolute value normalization. This ensures the maximal value of an individual galaxy's star formation history is 1, and minimum is at 0.

\begin{figure}
    \centering
    \includegraphics[width = \columnwidth]{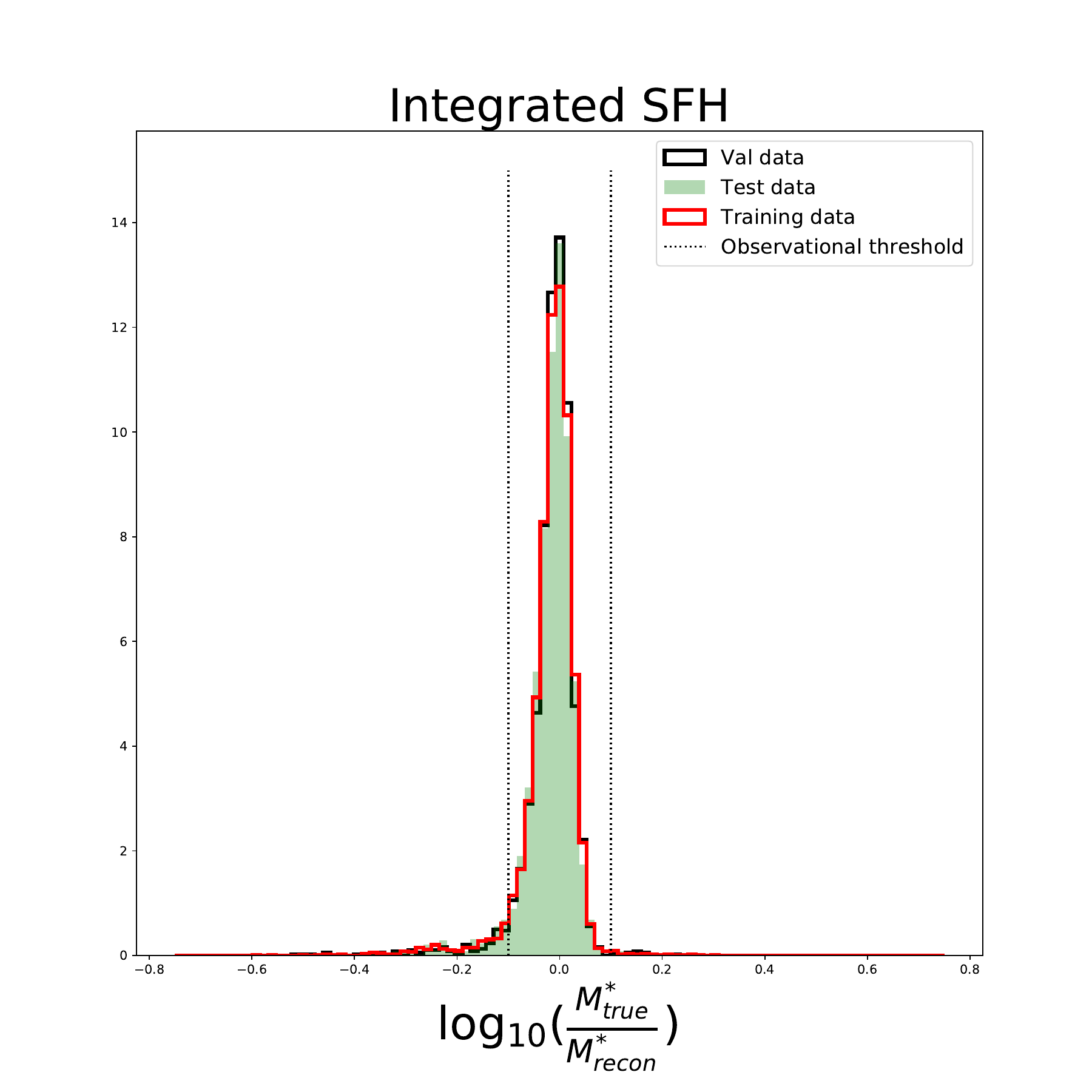}
    \caption{A normalized histogram of the integrated star formation histories, comparing between the original star formation histories and the reconstructed ones. Most ($99\%$) fall within the threshold of $\pm 0.1$ dex, with only a handful exceeding that, the greatest by no more than $\pm 0.21$ dex. The galaxy number does not correspond to the subhaloID.}
    \label{fig:recon_fig}
\end{figure}

 We determined the number of components needed by comparing NMF-reconstructed star formation histories with the originals. At around 4 components, we are able to sufficiently reconstruct most star formation histories (see Figure \ref{fig:recon_SFH}. Their integral gives a stellar mass within 0.1 dex of the original, in line with observational uncertainties (see Figures \ref{fig:recon_fig}). Specifically, we only find two galaxies with integrated stellar masses away from the original by 0.2 dex, and 47 away from the original by 0.1 dex. None exceed a deviation of 0.21 dex. This methodology was extended to metallicity histories, where we again found 4 components as being sufficient in representing the majority of metallicity histories. While we expect that a higher number of components would be needed to fit individual galaxies (e.g. \citealt{https://doi.org/10.48550/arxiv.2202.01809} require a stochastic burst in addition to 4 smooth NMF components in order to reproduce realistic galaxies), populations of galaxies can be described by a smaller number of smooth components \citep{Chaves_Montero_2021}.

Additional parameterizations for star-formation and metallicity histories were explored, e.g.: taking mass-weighted means of the metallicity, identifying the peak of the formation histories, comparing the timing of these peaks, and the net change in star formation history between adjacent bins. Ultimately, while incorporating these parameters did contribute to an interesting set of clusters, we found these clusters could be partially reproduced through using a smaller parameter space, populated by NMF representations of the star formation and metallicity histories. 

\subsection{Data extraction: Dark Matter Properties}\label{sec:massrates}

We use the halo mass assembly histories as computed in \citet{Chittenden_2022} and summarise how they are computed in this section.

The dark matter halo's mass accretion rate is defined using the Friends-of-Friends (FoF) group mass enclosed within a sphere whose density is 200 times the cosmic critical density at the given time, and whose radius is defined accordingly. We convert this to the rate of accretion onto the halo by finite differencing along the Main-progenitor-branch (MPB) with respect to the time $t_i$ at each snapshot, per Equation \ref{eq:Mhdot}.

\begin{equation}
\dot{M_h}(t_i) = \frac{M_h(t_i) - M_h(t_{i-1})}{t_i - t_{i-1}}
\label{eq:Mhdot}
\end{equation}

The amplitude and geometry of such accretion rates are useful indicators of the manner of ongoing halo formation. Smooth accretion of dark matter is usually characterised by small gradients in the accretion rate, whereas merger events present more instantaneous or sharply rising peaks.

\subsection{Defining clusters}

We considered and tested a number of clustering algorithms. We did this by clustering a subset of our parameter space; the peak star formation rate, mass weighted mean metallicity and stellar mass - a compressed representation of the parameter space we wanted to explore. We found that Gaussian Mixture Models were preferable, as the cluster labels they produced produced clusters with distinct populations, that aligned with the results of visual inspection, which is feasible in a low dimensional parameter space. Next, we had to decide on the number of clusters, as E-M algorithms like GMM require this to be dictated a priori.  Scoring metrics assign a score to a set of clusters, according to [\cite{scikit-learn}'s implementation of the Silhouette score \cite{rousseeuw1987silhouettes}, Calinski-Harabasz index \cite{calinski1974dendrite} and the Davies-Bouldin Index \cite{davies1979cluster}]. Although scoring metrics attempt to determine the optimal number of clusters, their accuracy is questionable- in our tests, different metrics often produce different results. Instead, we developed a "consensus framework", consisting of the following steps: 
\begin{itemize}

    \item  Clustering the entire dataset with $N=2,M_{BIC}$ clusters, where $M_{BIC}$ is the number of clusters as recommended by the Bayesian Information Criterion (BIC). BIC assumes a Gaussian distribution, and penalizes overfitting, therefore the number of components that extremizes the BIC is the upper limit, as it assumes fully Gaussian data 
\citet{2014statistics}.
    
    \item Scores are computed for different metrics ( Silhouette score , Calinski-Harabasz index, and the Davies-Bouldin Index). Each recommends a certain number of clusters.
    
    \item If the number of clusters suggested by all three metrics is the same, then there is a clear agreement on the clustering, and we run through it with the validation set to verify and use it as the definitive set of clusters.
    
    \item If the number of clusters is in agreement, but not in consensus (i.e. a 2-1 vote), we use the validity index, a scoring system defined by \citet{moulavi2014density}. If there is still no agreement, then it merits further visual inspection. A final decision is made based on visual inspection.

\end{itemize} 

We consistently found 3-4 clusters as being a well-supported choice. It also produced clusters whose parameter distribution appeared reasonable under visual inspection. The selection of three or four clusters is not wholly authoritative, as other clusterings could still be found to have distinct populations - the number of suggested clusters ranged from 2-5, depending on the score. 

We argue that our methodology supports the choice of 3 and 4 clusters, which we also confirmed by investigating if they are distinct via PCA.

 We ran PCA on each cluster and on the entire population of galaxies.  Comparing the eigenvectors in each case allows us to determine whether the sub-population being surveyed is distinct from the overall population.

\begin{figure*}
    \centering
    \includegraphics[width = \textwidth]{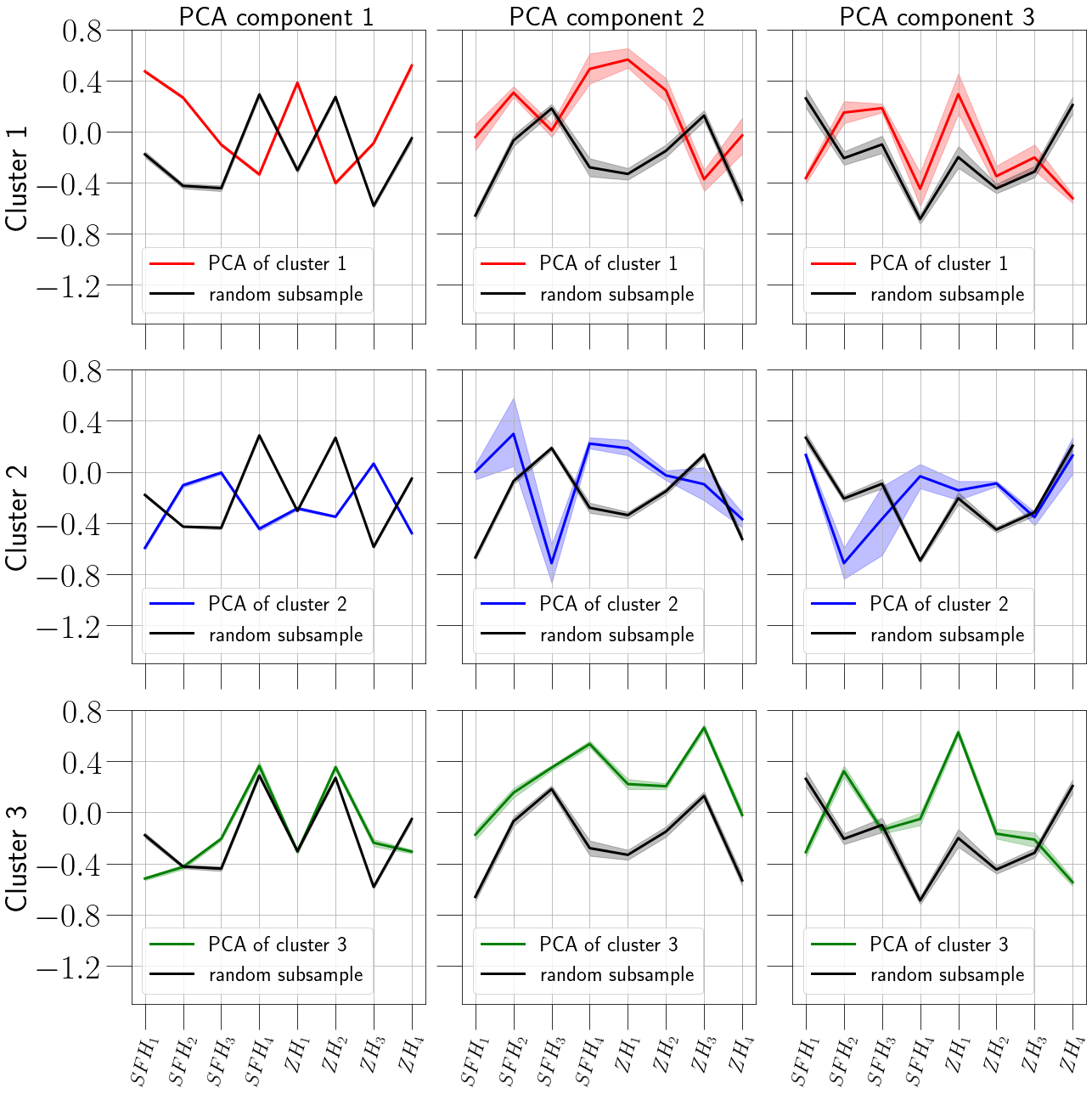}
    \caption{Comparison of the PCA  eigenvectors of both a given cluster (run 50,000 times) and a randomized subset of galaxies from IllustrisTNG. The shaded regions are the 5th/95th percentiles, and demonstrate that each cluster consists of populations whose PCA components do not fully overlap with each other, nor do they resemble the general population of 12535 galaxies used. This is how we establish that the subpopulations (as identified by our clusters) are distinct. The dark lines and their shaded regions are random equivalents of a selected population (i.e. a random subsample of galaxies, selected without the cluster labels was selected and run through the same procedure, to demonstrate that the populations identified differ from a random subsample).}
    \label{fig:distinct_ex_B}
\end{figure*}

Figure \ref{fig:distinct_ex_B} demonstrates this process. We first subdivide our dataset into four populations, each with a correspondent number of galaxies as in our clusters, but the galaxies themselves randomly selected. We then run a Monte-Carlo simulation of the PCA on these facsimilie subpopulations, recording their  eigenvectors. We then take  our clusters and run PCA on each of them, in the same manner (50,000 times for each subpopulations). We ran it 50,000 times for the randomly sampled subsets (black), and the clustered populations(colour) in  the resulting eigenvectors are plotted in Figure \ref{fig:distinct_ex_B}. Seeing how the distribution of both the random subsample's PCA and the correspondent subpopulation's PCA distributions vary, we can then conclude that these populations are considered "distinct". Although only shown here for one case, we perform an equivalent PCA analysis on all cluster sets detailed in the next section.

\subsection{Procedure}

Our analysis worked in two stages. First, we determined clusters with “observables” from IllustrisTNG. We defined sets of clusters based on the NMF compression of star-formation and metallicity histories and/or optical colour - see Table~\ref{tab:cases} for a summary. Second, we linked these clusters to dark matter halo parameters: the mass weighted age, the halo mass assembly history, the stellar in/ex-situ mass fractions, and the dark matter fraction.

 We also select for central and satellite galaxies by defining centrals as the most massive subhalo in a given group in the $FoF \ halos $ group catalog.  All other galaxies are treated as satellites. The breakdown of centrals and satellites is outlined in Table \ref{tab:satcentsplit}. We can then cluster the data and determine the cluster labels for the full data set and apply the central/satellite splits.  We evaluated the similarity of these distributions by also comparing the results of a two-sided KS test on the ex-situ mass fractions from mergers. This is because almost all the galaxies in our population have a non-zero ex-situ mass fraction from mergers, so comparing the KS test result for these statistic is critical. Our criterion for similarity was defined as the two  KS-test producing a p-value of greater than $0.05$. 
 
\section{Results}\label{sec:Results}

\begin{figure*}
    \centering
    \includegraphics[width =\textwidth]{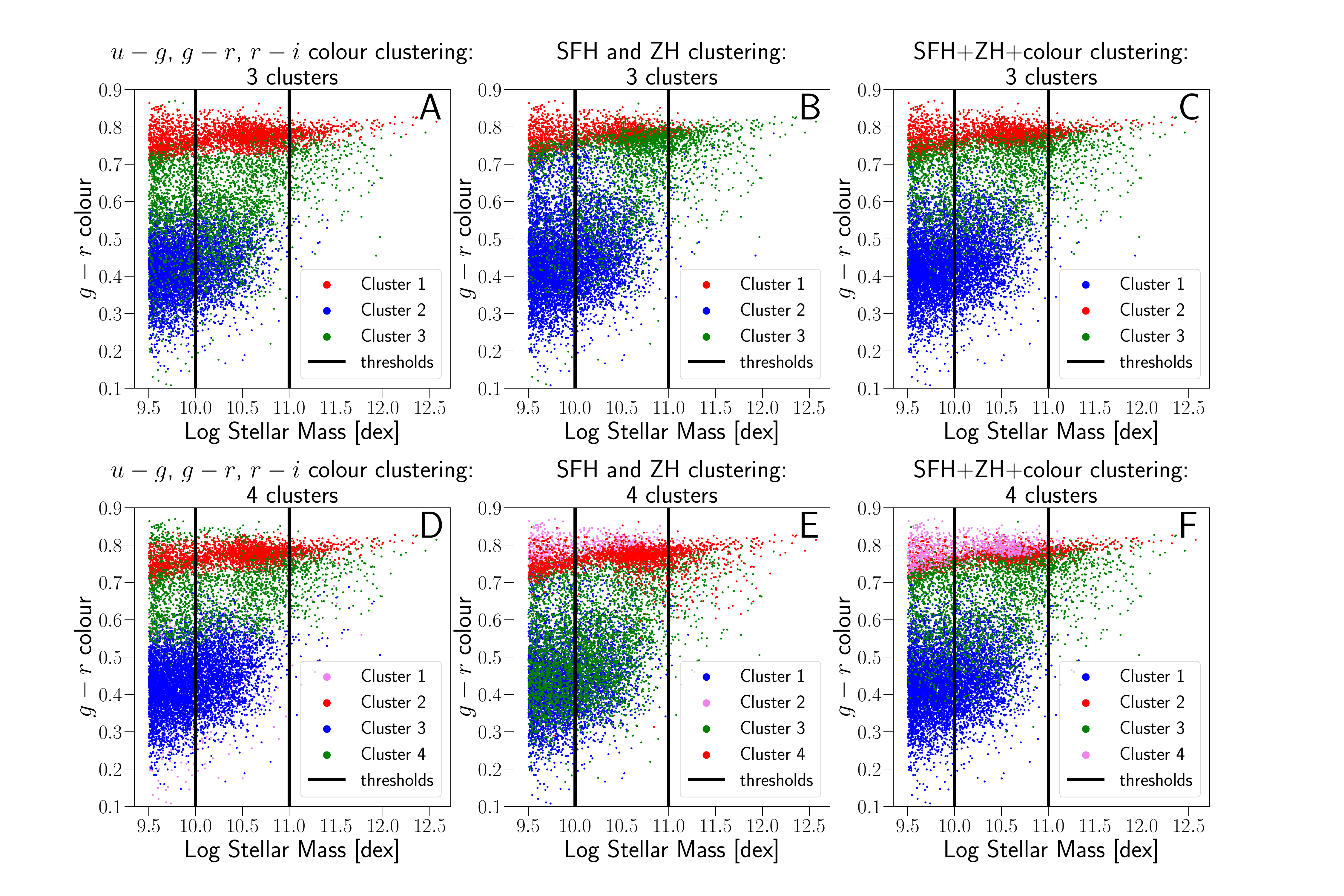}
    \caption{Colour-mass plot of the six different cases we explored in detail. This plot serves to situate the reader, covering the different clusterings we explored. These cases were the clusters that emerge when clustering  $u-g$,$g-r$,$r-i$ colours (left column), as well as NMF-4 component representations of the star formation and metallicity history (middle column), and the result of clustering the latter together (right column). The top row focuses on cases with three clusters, the bottom row focuses on cases with four clusters. Generally speaking, while there are a great deal of similarities between cases, this colour-mass plot only serves to situate the reader, as other metrics were needed to quantify the differences in these evolutionary histories. The black lines denote the mass range used for our fixed-mass figures. }
    \label{fig:Summary}
\end{figure*}

Figure \ref{fig:Summary} visualizes all of our cluster sets (as described in Table \ref{tab:cases}) as colour-stellar mass plots. Stellar mass was not used in the clustering, and optical g-r colour was only used in some of the sets. We show a colour-stellar mass diagram to situate the discussion, which we will often phrase in terms of a red sequence, a blue cloud or a green valley. We outline the general description of each case and cluster in Table \ref{tab:satcentsplit}, and \ref{tab:allcases} and  for a breakdown of their central/satellite membership.

For a more exhaustive overview of each section, we refer the reader to the subsections below. We ordered these subsections based on their importance to our task of identifying distinct populations.

%

Throughout our discussion of each set of clusters, we will refer to plots of mean star-formation histories, metallicity histories, halo assembly histories and histograms of the fraction of stellar mass assembled ex-situ by mergers and flybys. We diminish the role of stellar mass in our results by focusing our discussion - and our plots - in a fixed stellar mass interval, corresponding to 10 to 11 in $\log M_* / M\odot$. This interval samples most of our clusters well, whilst avoiding the extremes of the galaxy population. We have kept a consistent colour scheme in all plots, such that red represents the most quenched population, blue the most star-forming population, and green represents the remaining population. In the case of 4 clusters, we have represented the new population in pink.

We found that clusters determined using the star formation and metallicity histories were especially effective in identifying populations with distinct evolutionary histories. By contrast, clustering with broadband colours, while effective in differentiating between different star formation histories, was not as effective in distinguishing other evolutionary aspects. Here, we will highlight the different clusters we found, and summarize their features, both observable and non-observable, and how that distinguishes their respective evolutionary histories. We find that cases A, D with photometric colours produce only superficial distinctions in their populations as we discuss below. We also find that case C fails the two sided KS test, as the probability of two distributions being drawn from the same probability distribution function is significant ($p>0.05$). For case F, this complicates comparisons with case C, so we opt to study cases B and E in greater detail, and place them first in this section.

\begin{table}
\centering

\caption{This table shows the 
key features of each cluster set: their dimensionality (number of parameters clustered), the number of clusters produced, and a short description of the parameter space. }
\resizebox{\columnwidth}{!}{%
\begin{tabularx}{\linewidth}{c|ccX}
\hline
\textbf{Case} & \textbf{Number of clusters} & \textbf{Dimensions} &\textbf{Description} \\
\hline
A & 3 & 3 & Populations from clustering three photometric colours $u-g$, $g-r$, $r-i$. \\ \hline
B & 3 & 8 & Populations from clustering NMF compressions of star formation and metallicity histories. \\ \hline
C & 3 & 11 & Populations from clustering with parameter space of cases A and B.\\ \hline
D & 4 & 3 & Populations from clustering three photometric colours $u-g$, $g-r$, $r-i$. \\ \hline
E & 4 & 8 & Populations from clustering NMF compressions of star formation and metallicity histories. \\ \hline
F & 4 & 11 & Populations from clustering with parameter space of cases D and E.\\
\hline
\end{tabularx}%
}
\label{tab:cases}

\end{table}

 \begin{table}
     \centering
     \caption{Summary Table of each of the six cases explored in this work. Here, we enumerate the number of galaxies that were identified in a given case as central or satellite, and their breakdown across clusters. While there are some differences between our major cases of interest (cases B,C,E,F), we can see some analogous features. In particular, the satellite dominance of one of the clustered populations, typically inhabiting the lower red sequence is readily apparent for cases B and C.}
     \begin{tabularx}{\linewidth}{c|c|c|c|c|c}
          \textbf{Case} & \textbf{Cluster} & $N_{Centrals}$&$N_{Satellites}$ & $ \% central $&$\% sat $  \\\hline
          A&1&1043& 2048& 33.74 & 66.26 \\
          A&2&4523& 1628& 73.53 & 26.47  \\
          A&3&1671& 1622& 50.74 & 49.26 \\ \hline
          B&1& 390 & 1464 & 21.04 & 78.96  \\
          B&2&5200& 2696 & 65.86 & 34.14  \\
          B&3&1647 & 1138 & 59.14 & 40.86 \\ \hline
          C&1&5193&	2333&	69.00&	31.00\\
          C&2&760&	1723&	30.61&	69.39 \\
          C&3&1284&	1242&	50.83&	49.17\\ \hline
          D&1&179&	372&	32.49&	67.51\\
          D&2&930&	1830&	33.70&	66.30\\
          D&3&5121&	1879&	73.16&	26.84\\
          D&4&1007&	1217&	45.28&	54.72\\ \hline
          E&1&3369&	1737&	65.98&	34.02\\
          E&2&96&	701&	12.05&	87.95\\
          E&3&2804&	1664&	62.76&	37.24 \\
          E&4&968&	1196&	44.73&	55.27\\ \hline
          F&1&5022&	2259&	68.97&	31.03 \\
          F&2&714&	1091&	39.56&	60.44 \\
          F&3&1405&	1245&	53.02&	46.98\\
          F&4&96&	703&	12.02&	87.98\\
     \end{tabularx}
     \label{tab:satcentsplit}
 \end{table}

\begin{table*}
    \centering
    \caption{Summary table of the populations identified in Figure \ref{fig:Summary}, and the general descriptions of each of the clusters produced by our GMM. See section \ref{sec:Results} for a qualitative breakdown.}
    \resizebox{\textwidth}{!}{%
    \begin{tabularx}{\textwidth}{c|XXXX}
    \hline \textbf{Case}&\textbf{Cluster 1}&\textbf{Cluster 2}&\textbf{Cluster 3}&\textbf{Cluster 4}  \\ \hline
A& Red sequence population of galaxies: quiescent and high g-r colour &
  Blue cloud population of  galaxies: star-forming and younger &
  Green valley population of  galaxies. &  \\ \hline
B& Low mass red sequence population of galaxies mostly satellite dominated, with some overlap with cluster 3 in terms of mass, but with some drastic differences in its evolutionary history. & Blue cloud, primarily populated with star forming galaxies and some galaxies inhabiting the "green valley" &  High mass red sequence population, tapers off below stellar masses of 10.5 dex, with considerable overlap with cluster 1 of this population. Primarily dominated by centrals. & \\ \hline
C& Blue cloud population of star-forming galaxies. & Red sequence population of galaxies,  mostly quiescent. &
  Green valley population of star-forming galaxies. &
   \\ \hline
D& Sub-population of blue cloud population of galaxies, across various masses and parts of the colour gradient. &
  Red sequence population of galaxies: quiescent and high g-r colour & 
  Blue cloud population of  galaxies: star-forming and younger &
  Green valley population of galaxies in addition to some redder galaxies \\ \hline
E& Blue cloud population of galaxies.  Relatively active star forming population  &
  Very red subset of the red sequence. Predominantly lower mass galaxies up until 11 dex. &
  Green valley population, with some galaxies from upper red sequence. &
  Mostly red sequence galaxies ranging from low to high mass. \\ \hline 
F&Blue cloud population of galaxies generally a star forming population. &
  Portion of the red sequence across all masses, excludes less red portion. Relatively narrow range of g-r colour. &
  A segment of the red sequence, across all masses, notably excludes some redder, less massive galaxies. &
  Segment of lower red sequence typically a much redder population (in terms of g-r colour). \\ \hline
\end{tabularx}%
}
\label{tab:allcases}
\end{table*}

\subsection{NMF derived clusters- case B}
In terms of the mean star-formation and metallicity histories of each cluster are shown in Figure \ref{fig:SFHZH-B}, the dark matter halo assembly histories in Figure \ref{fig:MAH_caseB}, and the fraction of mass accreted by mergers and flybys in Figure \ref{fig:fMBplot}.

On the colour-mass diagram (Figure \ref{fig:Summary}), we found that the clusters split along the lines of a star forming, blue cloud population of galaxies (cluster 2), with two other clusters coming from a split in the "red sequence" population of quiescent galaxies, largely corresponding to their stellar mass,  but not entirely driven by it. In the mass range of 10-11 dex, one such sub-population (cluster 1) is satellite dominated ($78.96\%$)  with a late peak in its mass assembly history at approximately 6 Gyr in lookback time, while the other is primarily dominated by central galaxies with a consistently higher ex-situ mass fraction, sourced from mergers (cluster 3), see Figure \ref{fig:fMBplot} for reference. Moreover, the mass-weighted ages of the stellar matter and the dark matter halos can be taken in contrast with each other to highlight differences in the formation histories of these populations. In cluster 1, the halo mass weighted age is $7.68_{-1.33}^{+1.45} \ Gyr $, while its stellar mass weighted age is $9.91_{-1.29}^{+1.26} \ Gyr$, indicating that most of the star formation seems to precede significant growth of the final halo.  As convention, we denote the mass weighted quantities with their mean age and their 16th/84th percentiles.

\subsection{NMF derived clusters- case E}

By rerunning the clustering for four clusters, we can explore what additional sub-populations might be lurking in our dataset, and what features might distinguish them. The mean star-formation and metallicity histories of each cluster are shown in Figure \ref{fig:SFHZH-E}, the dark matter halo assembly histories in Figure \ref{fig:MAH_caseE}, and the fraction of mass accreted by mergers and flybys in Figure \ref{fig:fMEplot}. It is important to note that adding a cluster means that the dataset as a whole is re-clustered, meaning that some of the populations identified in a three-cluster case may not persist in a four cluster case.

Adding a fourth cluster produces a population that begins to significantly quench 8-6 Gyr ago, using Figure \ref{fig:SFHZH-E} as a reference. This also means we get a population quenching 5-2 Gyr ago, and another just beginning to quench around 0.5-1 Gyr ago. The final population best fits the characteristics of a blue cloud population: star formation is continuing relatively unabated.

This subpopulation that quenched long before the others is extremely satellite dominated ($87.95\%$) compared to the other populations, but also a relatively small subpopulation (797 galaxies out of a set of 12535, see Table \ref{tab:allcases}). It has the greatest difference in median ex-situ mass fraction for mergers (half an order of magnitude), and second largest for flybys, indicative that this is a population where mergers play a critical role in its evolution, and is more akin to other populations that are merger driven, i.e. cluster 4. What is notable as well, considering clusters 1,2,3 is that while the mean star formation histories all peak at similar times, their quenching timescale as a population varies greatly. Moreover, we find in terms of mass weighted ages for the halo and stellar mass, only clusters 1 and 3 have an older mass weighted stellar age than halo age, with a difference of over 3 Gyr for cluster 1. This seems to indicate that most of the stars have formed prior to the final halo being fully assembled. By contrast, a blue cloud population has significantly younger stars than its halo’s mass weighted age, indicative of the opposite. Given there is also a distinct lack of overlap between the two quenched populations (clusters 1 and 3), per Figure \ref{fig:SFHZH-E}, it is interesting that these populations have galaxies peaking at similar times of their star formation history, but quenching across a wide range of time scales.

When we end up increasing the number of clusters to 4, we found that this new population ends up selecting for galaxies that are relatively lower in mass, but are very extinct, and very much dominated by satellites.

As in the other populations, the mass weighted age of the halo is older than the stellar matter. In the high ex-situ mass fraction population, this amounts to ages of $8.68_{-0.70}^{+0.71} \ Gyr$ and $8.27_{-1.15}^{+1.19} \ Gyr$, ages that practically overlap, indicating that these mergers would play a role in star formation and evolution. 

In focusing on the non-observable quantities, we find that the distribution of satellite galaxies in these two populations to be of interest, as the older population (cluster 1) is distinctly dominated by satellites, and with a smaller flyby and merger ex-situ fraction than cluster 3, who is split more evenly in terms of their central/satellite demographics. This demonstrates that the addition of a single cluster reveals interesting populations with characteristic features, namely, the sheer diversity of quiescent galaxies, possibly linked with different evolutionary histories. This is dissected in Section \ref{sec:Discussion}.

\subsection{Colour-based clusters - case A}

In case A, we cluster based on optical colour only, and consider three clusters. We found that while there were superficial differences seen in the colour-mass diagram (Figure \ref{fig:Summary}), these did not extend to differences in the stellar evolution or more importantly, the ex-situ mass fractions and the mass assembly history. The metallicity histories, unlike the star formation histories overlap significantly . The ex-situ merger mass fraction medians between the blue cloud of cluster 2 and the green valley of cluster 3 are (-1.06, -1.11 vs -1.09, -1.21  for central and satellite subpopulations respectively), and -3.02,-3.18 vs -2.99,-3.20 for flyby mass fractions. The overall distributions are very similar as well, underscoring that these populations are not really as distinct as the rest. This distinction criterion is ultimately qualitative.

In contrast to the populations discussed in the previous subsections, here we highlight the limitations in distinguishing between galaxies across parameters ranging from observables (metallicity histories) to intrinsic parameters (dark matter halo mass assembly history and merger histories).

\subsection{Colour-based clusters - case D}

In case D, we cluster based on optical colour only, but now consider four clusters. The mean star-formation and metallicity histories of each cluster are shown in Figure \ref{fig:SFHZH-D}.

While clustering with only photometric colours does not really allow us to uncover much about the dark matter halo's evolutionary history, the four cluster case produced novel populations with distinct features that are worth examining in an observational context. In the discussion, we will opine on some possible characteristics, but here we will summarize the key attributes. The clusters in this case are very similar to case A's, with a star forming population, a quiescent one, and a population that inhabits in between (green valley). The added caveat is that this latter population incorporates some lower mass and redder galaxies, and that the last cluster (cluster 3), incorporates a subset of galaxies from the "blue cloud". What is apparent with cluster 1's galaxies is that while these galaxies at fixed masses have similar star formation and metallicity histories, they tend to slightly exceed cluster 3's average in star formation history and metallicity history in the last 2 Gyr (Figure \ref{fig:SFHZH-D}). Their ex-situ mass fraction distribution is unique in that the merger mass fraction has a similar median as cluster 1, but the flyby mass fraction is considerably higher, median value of -2.45 for satellites and -3.08 for centrals, indicative that unlike any of the populations found in this clustering or cases B,E, the satellites are sourcing more of their ex-situ mass fraction from flybys than centrals, and the distribution is almost bi-modal. This must be taken with the caveat of relatively low statistics, with only 551 galaxies total in the population and it being ($67.51\%$) satellite dominated, versus $66.30\%$ for the red sequence and $26.84\%$ for the blue cloud population. We discuss the potential this population has as a shorthand for identifying post-starburst galaxy
(PSB) populations in our discussion.

\subsection{Colour and NMF derived clusters - case C}

We explored the clusters that would emerge if we combined the parameter spaces used in previous sections: NMF representations of star formation and metallicity histories alongside photometric colours. Our reasoning was twofold: to explore the overall utility of photometric colours as additional parameters, and to highlight additional sub-populations of interest. 

Despite the populations produced greatly resembling those found in case A in the colour-mass diagram, we find that unlike case A, these populations have distinct metallicity and star formation histories - demonstrating that the clustering adroitly differentiates between different star formation and metallicity history profiles. Notably, the green valley population (cluster 3) quenches along a distinct timescale compared to the redder galaxies of cluster 2 (starts quenching around 1 Gyr lookback time,versus 6.5 Gyr lookback time). This distinction is also readily apparent in the metallicity history, which has each population peaking at different times for its metallicity history, demonstrating the diversity of each population's chemical enrichment. We can also extend this analysis to non-observable parameters, which were a source of contention in using the clusters produced by case A . For the ex-situ mass fractions , we find that the addition of NMF components to our colour clusters does noticeably alter the composition. However, these changes do still come with some of the issues initially raised. The largest change is seen in the distributions of the flyby and merger mass fractions of galaxies in the green-valley, cluster 3.

Here, we find that the two-sided KS test, gives its only significant ($p>0.05$), with a $p=0.2$ for overlap between cluster 2 and 3. This underscores existing doubts we had about this population, as it shows that there is a significant probability of these two populations being the same. The decision regarding distinct populations is ultimately qualitative, but we use this as grounds to favour case B over case C as being a much more distinct population, which is the ultimate goal of this paper: distinct populations.

\subsection{ Colour and NMF derived clusters - case F}

For the final case, we explore the addition of a fourth cluster to the parameter space used in case C.

We find that this new cluster produces a population that is extremely satellite-dominated, and occupies a low mass, but very quiescent branch of the red sequence. This is underscored by the distinctly sharp peaks in its star formation and metallicity history, and rapid quenching of this population, with its mean having quenched by 6-7 Gyr lookback time.

We find that cluster 2 and 3 have strongly overlapping medians for their ex-situ merger mass fractions (-0.82, -1.12 for centrals and satellites, vs. -0.99,-1.21), with the overall distribution of flyby mass fractions between the two populations being similar. For cluster 3 the distribution of log flyby mass fraction is overall bimodal, as in case C with peaks at approximately -3.5 and -1.5 . There is a larger proportion of central galaxies at lower merger mass fractions than in the red sequence population, indicative of a broader distribution. Compared to case E, cluster 2 and 3 are qualitatively too similar.

\begin{figure}
    \centering
    \includegraphics[width = \columnwidth]{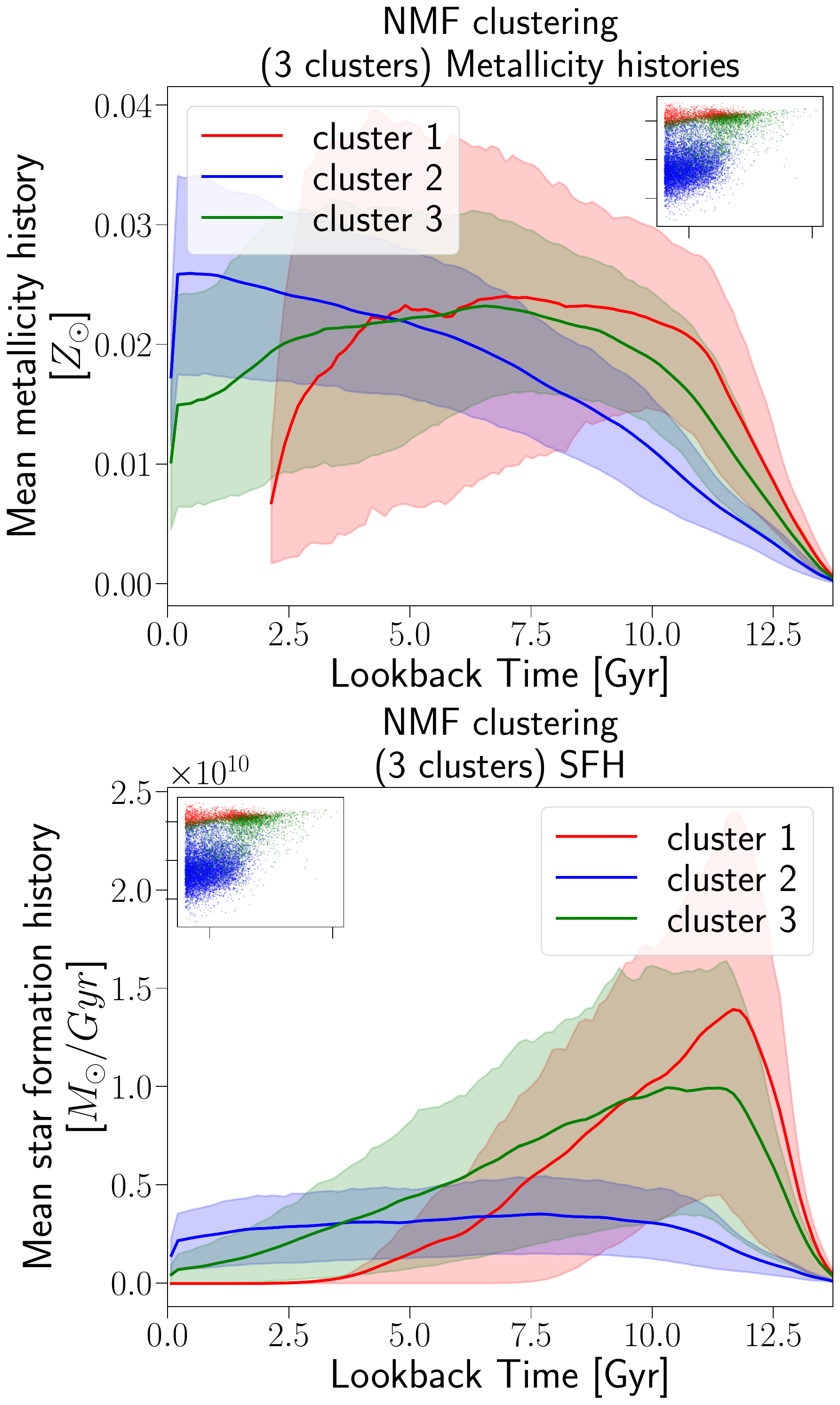}
    \caption{ \textbf{case B:} clusters generated using smoothed NMF components of the star formation and metallicity histories. The ensuing populations are distinct in terms of both metallicity histories and star formation histories, with the most striking feature being a division of the red sequence, with the upper portion (higher mass) having yet to fully quench, even along a fixed mass interval. Galaxies in this upper population (cluster 3) clearly seem to be experiencing something that prevents them from quenching as quickly as those seen in cluster 1. See Discussion for more details.}
    \label{fig:SFHZH-B}
\end{figure}
\begin{figure}
    \centering
    \includegraphics[width=\columnwidth]{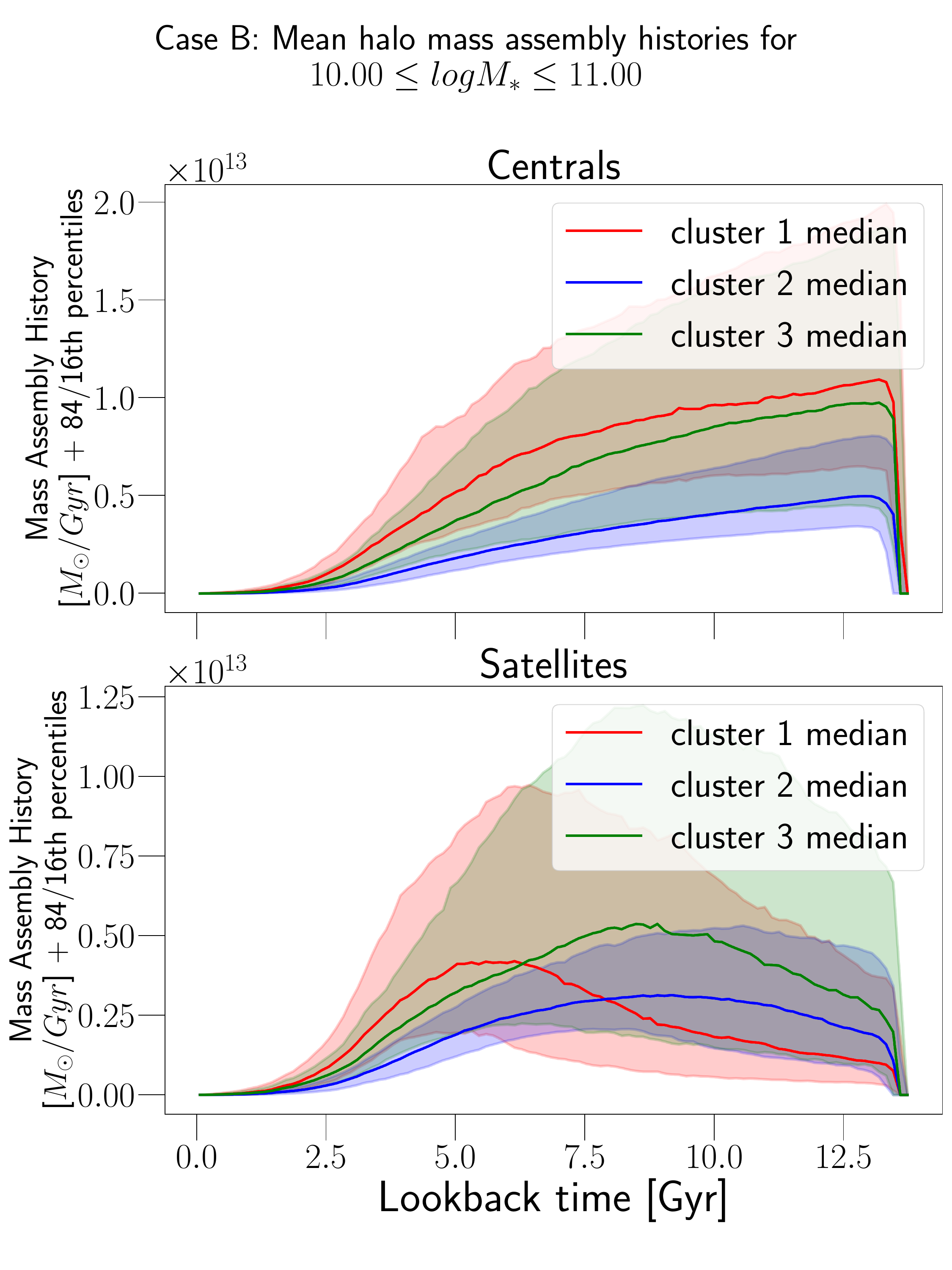}
    \caption{\textbf{case B:}Median mass assembly histories of the dark matter halos of the galaxies from case B, with 1$\sigma$ values shaded in. Here, we find that like all other cases, the greatest differences can be seen in the halo mass assembly history of the satellite galaxies of a given cluster, rather than their centrals. We find that besides a slight variation in the overlap of cluster 1 and 3's median histories for their centrals, the histories of the central galaxies barely vary, peaking early and diminishing as we approach the present. By contrast, the satellites in each of these clusters have distinct periods of time at which their mass assembly histories peak (lower panel). The satellite dominated population of cluster 1 peaks the most recently, approximately 6 Gyr ago. Meanwhile, merger-rich cluster 3 peaks further in the past, at 8 Gyr lookback time. These peaks might be suggestive of a period of more active mergers occurring for these respect galaxies, per Figure \ref{fig:SFHZH-B} at these times. It is also demonstrative of the underlying differences identified by our clustering. Note that the solid line is  a median.  }
    \label{fig:MAH_caseB}
\end{figure}

\begin{figure}
    \centering
    \includegraphics[width=\columnwidth]{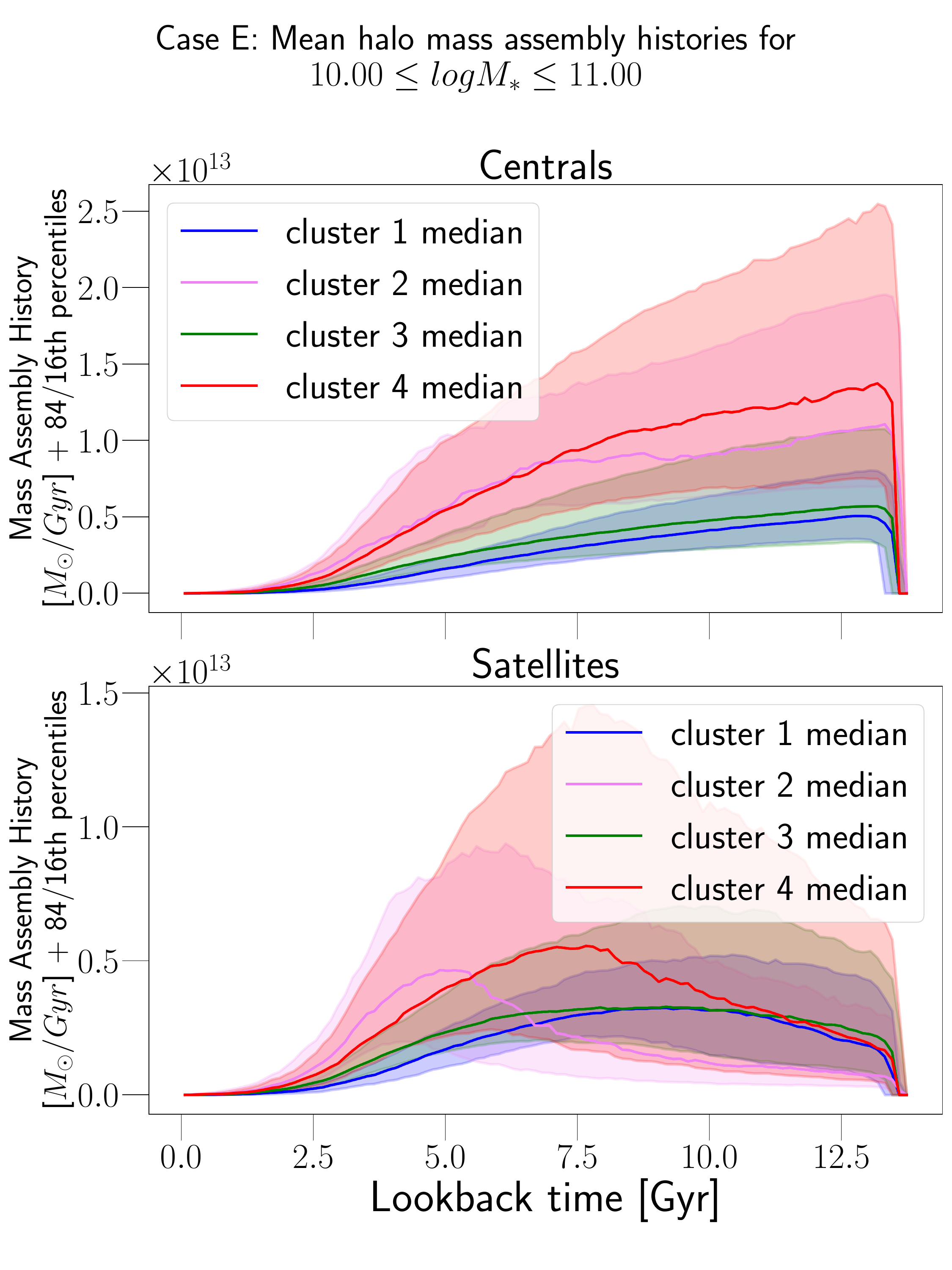}
    \caption{\textbf{case E:}Median mass assembly histories of the dark matter halos of the galaxies from case E. Here, we find that like all other cases, the greatest differences can be seen in the halo mass assembly history of the satellite galaxies of a given cluster, rather than their centrals. In fixing for stellar mass, we find that the satellites in cluster 2 have the most recent peak in their assembly history around 5 Gyr ago, cluster 4 with a peak around 8 Gyr ago, and clusters 1 and 3 having overlapping peaks around 10-11 Gyr ago. A point of contrast is comparing these recent peaks in the median mass assembly history with the star formation histories in Figure \ref{fig:SFHZH-E}, which are indicator of the role this growth in the halo plays in the stellar evolution. We explore this in detail in the discussion.}
    \label{fig:MAH_caseE}
\end{figure}

\begin{figure}
    \centering
    \includegraphics[width=\columnwidth]{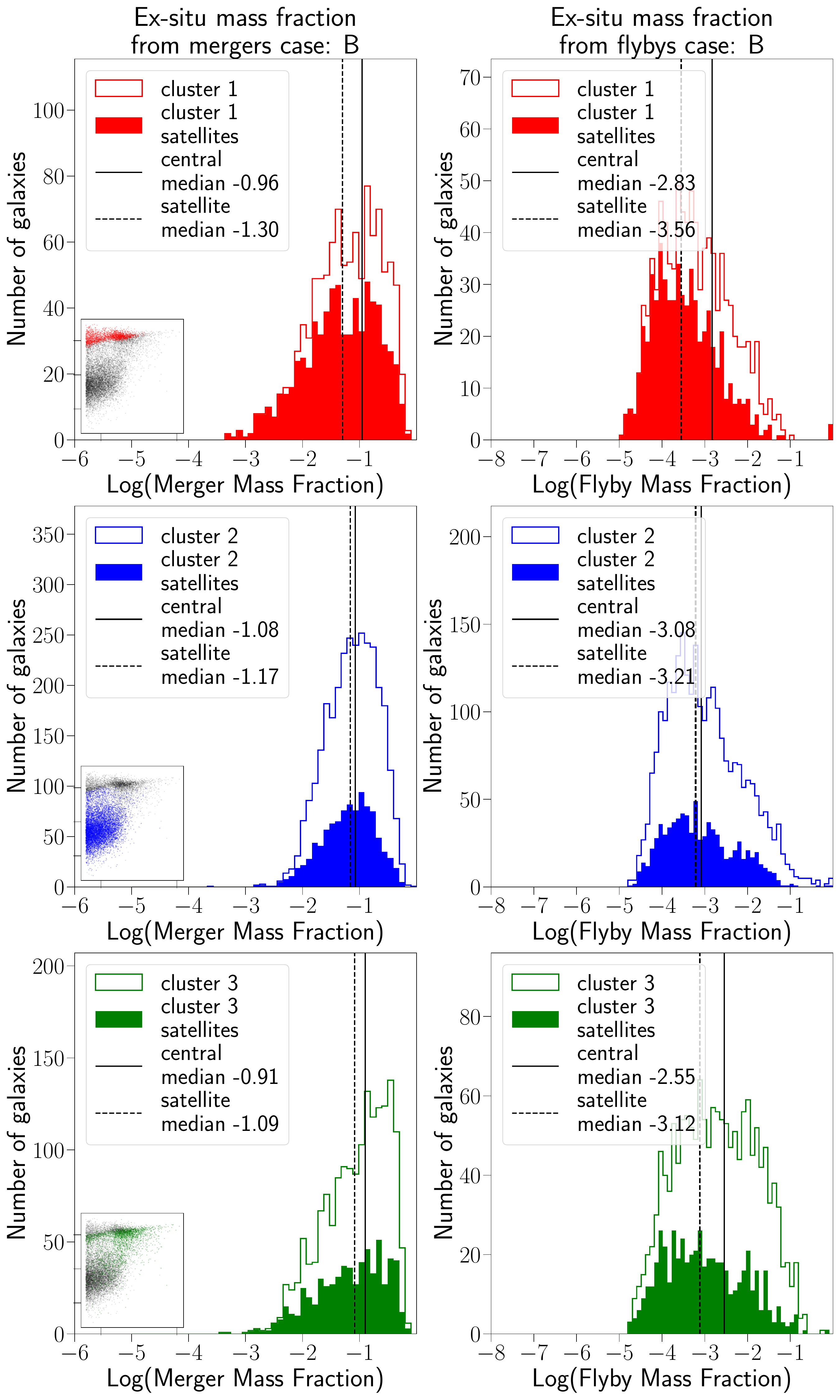}
    \caption{\textbf{case B:}Distribution of ex-situ mass fraction from mergers (left column) and flyby events/ongoing mergers (right column) from IllustrisTNG-100 galaxies, split by cluster membership. Both the centrals and satellites in quiescent populations like clusters 1 and 3 for case B have very distinct peaks in the merger mass fraction and flyby/ongoing merger mass fraction distributions, that are distinct between centrals and satellites. Notably, it seems that galaxies in cluster 3 of case B, especially central galaxies, are likely to source an order of magnitude or more of their ex-situ mass fraction from mergers than galaxies in the other sub-populations. Considering this in conjunction with the mean star formation histories of figure \ref{fig:SFHZH-B}, this suggests at the role mergers might have in the rate of quenching of these galaxies. While these distributions do shift when fixing for mass, the distributions are still distinct.}
    \label{fig:fMBplot}
\end{figure}

\begin{figure}
    \centering
    \includegraphics[width = \columnwidth]{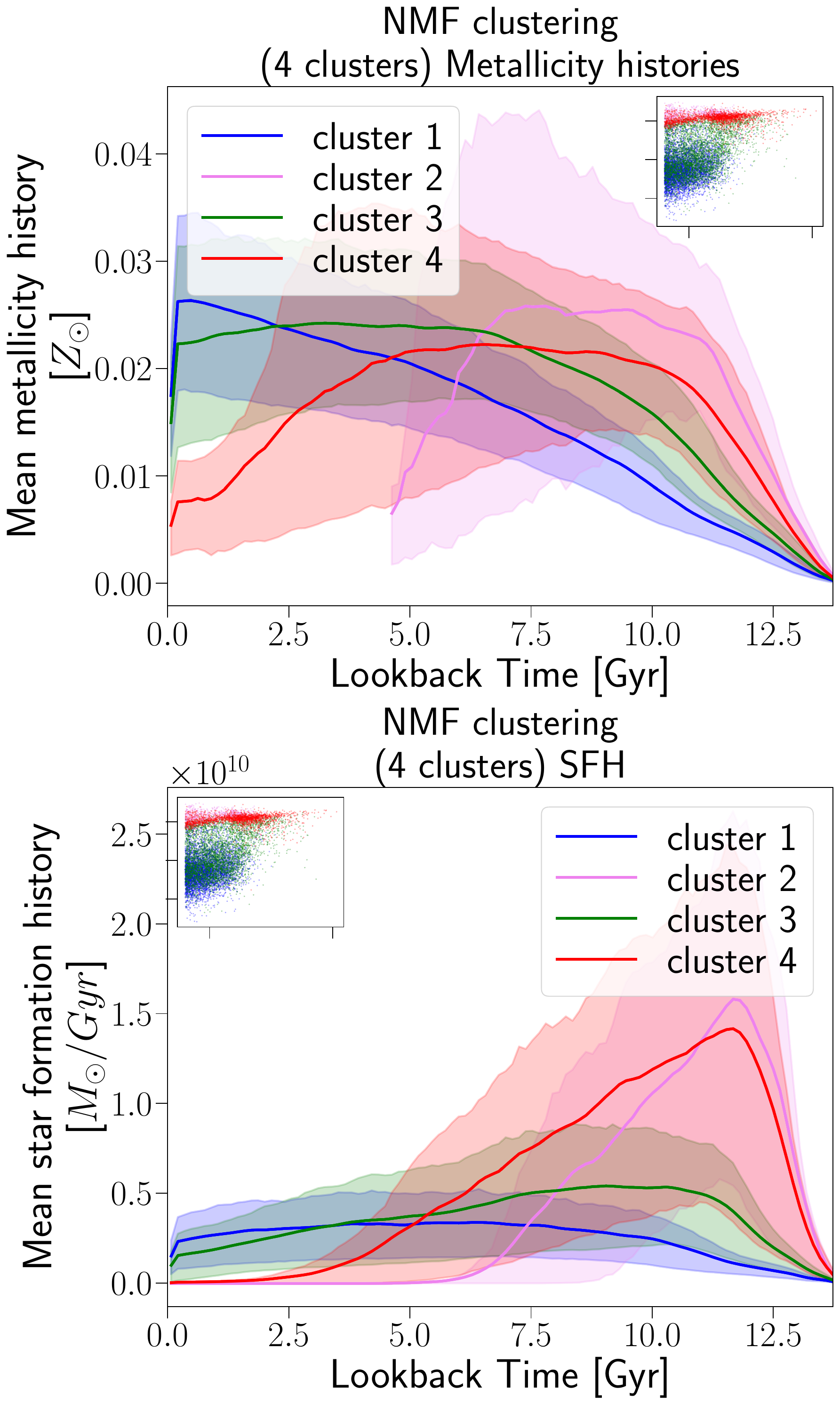}
    \caption{ \textbf{case E:} Plot of mean star formation history and metallicity histories with 1$\sigma$ regions shaded in, with clusters generated using smoothed NMF components of the star formation and metallicity histories. The ensuing populations are distinct in terms of both metallicity histories and star formation histories. The most striking feature is the split of these means: three out of the four populations are at some stage of quenching, with two of them having decisively quenched (clusters 2,4).  }
    \label{fig:SFHZH-E}
\end{figure}

\begin{figure}
    \centering
    \includegraphics[width = \columnwidth]{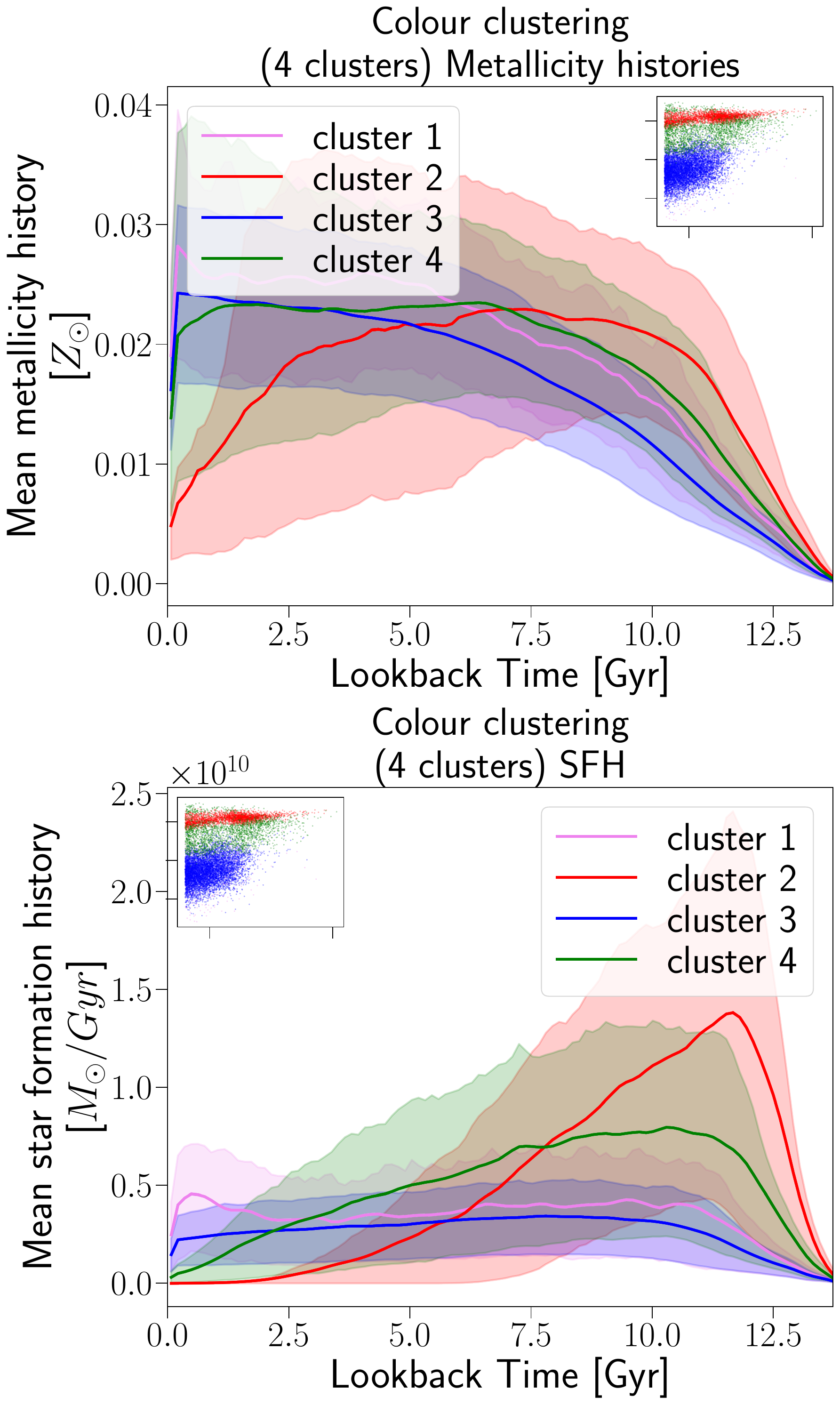}
    \caption{ \textbf{case D:}Plot of mean star formation history and metallicity histories with 1$\sigma$ regions shaded in, with clusters generated using photometric colours. The ensuing populations are not as distinct as in other cases discussed (e.g. Figures \ref{fig:SFHZH-B}-\ref{fig:SFHZH-E}, but special attention should be paid to cluster 1's late evolution in the last few Gyr. As we argue in the Discussion, this could serve as a shorthand for identifying post-starburst galaxies. These galaxies possess: an interesting evolutionary history, as evidenced by its ex-situ mass fractions.}
    \label{fig:SFHZH-D}
\end{figure}

\begin{figure}
    \centering
    \includegraphics[width=\columnwidth]{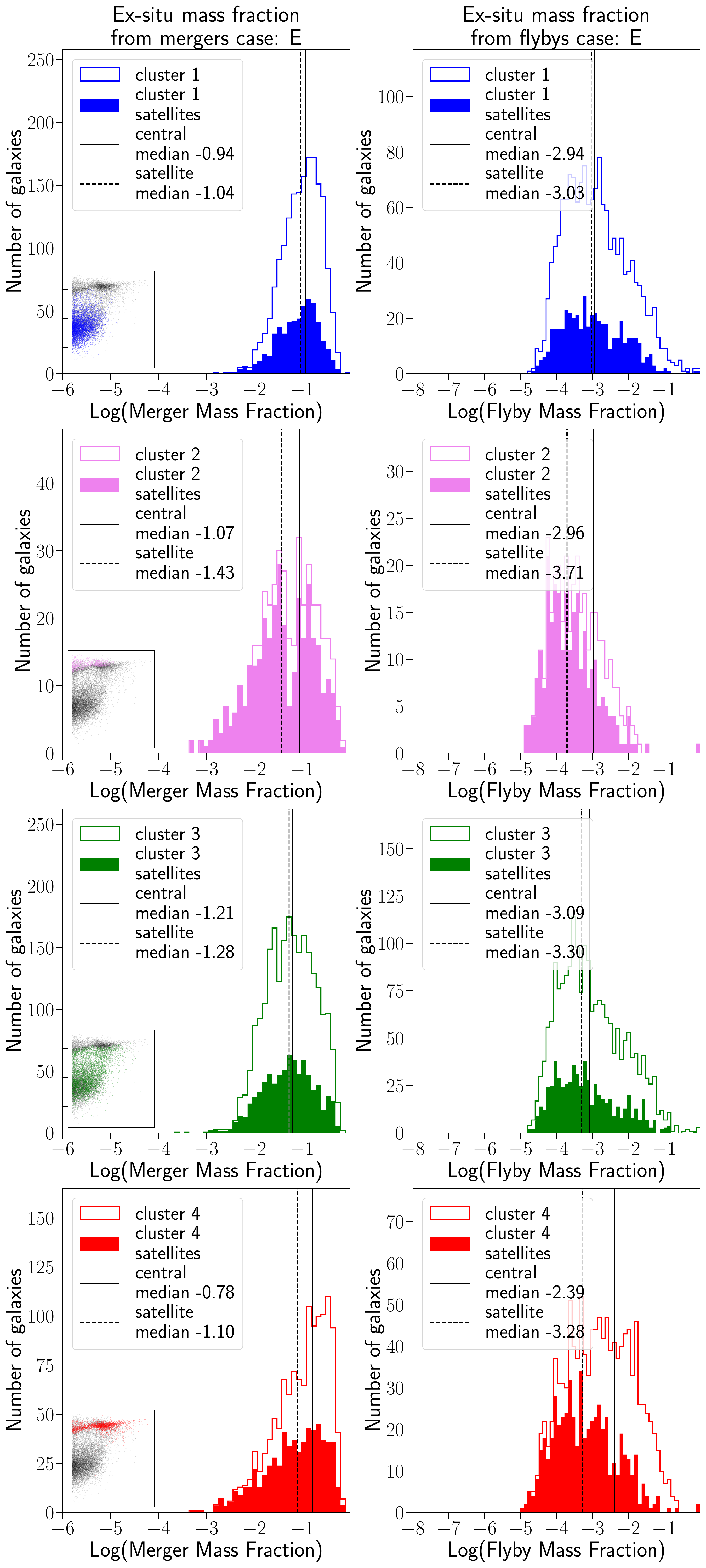}
    \caption{\textbf{case E:}Distribution of log ex-situ mass fraction from mergers $f_M$ (left column) and flyby events $f_{Flyby}$ (right column) from IllustrisTNG-100 galaxies, split by cluster membership. Cluster 4 is notable in that the distribution of $f_{Flyby}$ across centrals and satellites different medians and peak around -2 and -3 respectively -- a stark contrast to other quenched populations (Clusters 2 and 3), whose $f_{Flyby}$ is not as decoupled. This indicates mergers might play a role in delaying quenching in cluster 4 compared to these other populations. Of interest is cluster 3, whose wide spread in the distribution of $f_M$ might be indicative of differences in the merger history compared to other clusters. One could posit it challenges whether or not this population is actually a green-valley population, as mergers are expected to be a key component in the transition from blue to red populations.}
    \label{fig:fMEplot}
\end{figure}

\section{Discussion} \label{sec:Discussion}

\subsection{The effectiveness of using photometric colours vs NMF }

While star formation histories are an important marker of stellar and baryonic evolution within galaxies, as we show, they do not tell us everything. Cases A and D demonstrate the failure of colour-based clusters to produce a clear distinction beyond the star-formation histories. Examining the mean star formation histories reveals an overlap between populations, and looking at the mass weighted ages of the stellar and halo components, we find that for case A, stellar mass weighted ages of $9.48^{+1.41}_{-1.32} \ Gyr $, $6.24^{+0.87}_{-0.84} \ Gyr$, $ 7.01^{+1.01}_{-0.97} \ Gyr$ for each of its clusters, and mass weighted mean metallicities of $0.021^{+0.0022}_{-0.0027} \ Z_{\odot}$, $0.017^{+0.0022}_{-0.0024} \ Z_{\odot} $, $0.019^{+0.0016}_{-0.0021} \ Z_{\odot}$. Similarly for case D: $6.08^{+1.19}_{-1.06} \ Gyr $, $9.55^{+1.34}_{-1.27} \ Gyr$, $ 7.01^{+0.86}_{-0.84} \ Gyr$ , $ 7.72^{+1.08}_{-1.10} \ Gyr$ for each of its clusters, and mass weighted mean metallicities of $0.022^{+0.0011}_{-0.0046} \ Z_{\odot}$, $0.020^{+0.0022}_{-0.0023} \ Z_{\odot} $, $0.017^{+0.0024}_{-0.0025} \ Z_{\odot}$ , $0.020^{+0.0018}_{-0.0030} \ Z_{\odot}$.


When we consider non-observables we consistently see overlap in the ex-situ mass fractions and median mass assembly histories in the photometry-only clusters, further reinforcing that the divisions produced here do not capture the differences in evolutionary histories we are looking for and find in other clusters.  Particularly, looking at the ex-situ mass fraction for case A, between cluster 2 and cluster 3, we find that the overall distributions for both of these populations peak at very similar log mass fractions, with overlapping medians for flyby and merger mass fractions (-1.06,-1.11 vs -1.09,-1.21 for mergers, -3.02,-3.18 vs -2.99,-3.20 for flybys). This is even more pronounced in case D, where clusters 1 and 3 have overlapping merger mass fractions, with overlapping medians for both the centrals and satellites in these populations ( -0.88,-1.20 for cluster 1, -0.83,-1.18 for cluster 3).  

One population that is of lingering interest, and that we address later, is cluster 1 of case D. This is a subpopulation of galaxies whose star formation history and position on the $M_{*} - (G-R)$ colour plot greatly resembles the “blue cloud” galaxies identified in cluster 2, but is a satellite dominated population, with a significantly larger ex-situ mass fraction from flybys than cluster 2, particularly from its satellites (median value of -2.45 versus -3.27). In addition, the mean star formation and metallicity histories for this population exhibit a notable peak in the last 2 Gyr of lookback time - indicative of the presence of stars of that age. This population is exclusive to this clustering (4 clusters, using broadband colours only). Partly this will be driven by the fact that our NMF basis does not have the resolution at young ages to reliably pick up this population. Colours, on the other hand, are particularly sensitive to the SFH in the last Gyr \citep{Chaves_Montero_2020}, and can be combined effectively to identify post-starburst galaxies \citep{Wild_2014}. The sensitivity of optical broadband colours to recent star-formation combined with the heavy compression of SFHs into four NMF basis (with poor resolution at late times) likely explains why we can pick up this population more easily in case D. As we discuss later, this population can be linked to post-starburst galaxies identified elsewhere in the literature.

In contrast to these cases, cases B and E host significant differences in their mean star formation and metallicity histories, see Figures \ref{fig:SFHZH-B},\ref{fig:SFHZH-E}. This also manifests in the mass weighted ages of the stellar and halo components: $9.91_{-1.21}^{+1.26} \ Gyr$ for cluster 1,$6.26_{-0.78}^{+0.82} \ Gyr$ for cluster 2, $8.27_{-1.15}^{+1.19}  \ Gyr$for cluster 3 and $7.68_{-1.33}^{+1.45} \ Gyr $ for cluster 1, $9.09_{-0.55}^{+0.64} \ Gyr $ for cluster 2 and $8.88_{-0.72}^{+0.70} \ Gyr $ for cluster 3 of case B respectively.

$5.81_{-0.63}^{+0.65} \ Gyr$ for cluster 1,$10.59_{-0.86}^{+0.92} \ Gyr$ for cluster 2, $7.22_{-0.75}^{+0.76}  \ Gyr$ for cluster 3 and, $9.32_{-1.09}^{+1.22} \ Gyr $ for cluster 4.  $9.19_{-0.53}^{+0.65} \ Gyr $ for cluster 1,  $7.08_{-0.72}^{+0.70} \ Gyr $ for cluster 2,  $8.93_{-0.57}^{+0.57} \ Gyr $ for cluster 3 and $8.48_{-1.12}^{+0.97} \ Gyr$ for cluster 4 of case E respectively.Note how the satellite dominated population of cluster 1 has, on average, a younger halo than its stars. This is the first of many indicators about how the differences uncovered extend beyond the parameters we clustered.

\subsection{Analysing non-observables to distinguish between evolutionary histories }
Non-observable parameters such as the merger history, dark matter halo mass assembly history, ex-situ mass fractions, and  mean mass weighted ages reveals important information regarding the demographics of galaxies in the populations of the cases we identified, and how they behave.

In this section, we will focus on analysing in detail the populations revealed in cases B and E. The results with colour clustering (cases A,D) have shown while clustering with photometric colours does yield some superficial differences in the populations identified, the most interesting were found with clusterings that made use of star formation and metallicity histories. Our preference of cases B, E over C, F stems from case C having one population where it had a significant ($p>0.05$) p-value from a two-sided KS-test, meaning that the distributions studied (ex-situ merger mass fraction) of two populations had a very high probability of being similar.

The populations identified in case B,E are distinct, and highlight at how the existence of a green valley is not that well supported, as splitting along the red sequence, roughly based on the central-satellite status of a galaxy, reveals much more substantial information about its evolutionary history, as corroborated by significant differences in the ex-situ mass fraction (mergers) and mass assembly history. 

\subsubsection{Case B: Three cluster populations}
Case B results in a single star forming population identified, cluster 2, the remaining two splitting along the red sequence. When we keep stellar mass fixed, we find significant differences in their mean star formation histories and metallicity histories. Specifically, while both peak at a similar time, the time it takes for the mean SFH to drop to zero differs greatly (Figure \ref{fig:SFHZH-B}). This is supported by a difference seen in the mass assembly history and ex-situ mass fractions -  where the redder, more satellite dominated population (78.96\%) of cluster 1 has a similar distribution of ex-situ mass fraction between centrals and satellites, cluster 3, the higher mass end of the red sequence, has centrals with a higher ex-situ mass fraction, its distribution peaking at a larger fraction (Figure \ref{fig:fMBplot}). This indicates that these centrals are sourcing a greater proportion of their ex-situ mass from mergers and that mergers in cluster 3 seem to allow star formation to continue for so long compared to galaxies of a similar mass in cluster 1.

The median mass assembly histories complicate this picture, particularly when comparing the satellite subpopulations, which seem to have a “late peak”, coinciding at around 6 Gyr ago for cluster 1, 8-9 Gyr for cluster 3 (Figure \ref{fig:MAH_caseB}). This means that a major change happened in cluster 1, often after star formation stopped, while in cluster 3, this seems to have preceded quenching. The halo’s mass weighted age is $7.68_{-1.33}^{+1.45} \ Gyr$, the stellar mass weighted age is $9.91_{-1.20}^{+1.25} \ Gyr$ in cluster 1 vs $8.88_{-0.71}^{+0.70}$ and $8.27_{-1.15}^{+1.19} \ Gyr$ respectively for cluster 3. This signals that most of the stellar mass formed long before the halo gained all its mass. This is contrary to other populations, and when contrasted with cluster 3, seems to indicate how mergers might have allowed the quenching to be delayed and forestalled with mergers or how mergers brought in gas or younger stellar populations. So clearly, the satellite dominated case has older stars on average than the halo it inhabits, but centrally dominated populations do not. The blue cloud population has a larger gap in its age, like cluster 1, but cluster 3, despite inhabiting the red sequence like cluster 1, has an overlapping range of ages, with the stars being a little younger than the halo. 

This could corroborate the peak seen in the MAH of cluster 1 (Figure \ref{fig:MAH_caseB}) but raises significant questions about the merger history of these populations, which could be explored in a later work.

\subsubsection{Case E: Four cluster population}

In case E, we aimed to explore how the clustering changes when we re-cluster the population and allow for four populations instead of three. 

Interesting information can be gleaned when we examine indicators of the halo’s evolution and merger histories. The median mass assembly history of these satellite galaxies of these three populations(clusters 2,3,4)  peak at 5 Gyr, 9 Gyr and 8 Gyr ago in lookback time, approximately. By 5 Gyr ago, cluster 2 is almost entirely quenched, making it appear to be a more exaggerated version of cluster 1 from case B – satellite dominated, visible shift in halo assembly history after quenching. Meanwhile, the peak in the other two suggests, that much like what was discussed in case B, something is a bit different with satellite galaxies here. 

Extending to ex-situ mass fractions, we find that cluster 4 is the population with a distribution of ex-situ mass fraction that is significantly greater than the rest, both in overall distribution and its merger mass fraction medians (-1.10,-0.78 for satellite and central). What is even more notable is the order of magnitude difference seen in cluster 2 and cluster 4’s ex-situ mass fraction from flybys, between centrals and satellites, with centrals in both cases having a distribution peaking at a higher mass fraction than satellites. This indicates that the centrals in both cases are evolving slightly differently when compared to cluster 3. Mergers clearly play a critical role in these populations. With such a critical difference seen in the flyby fraction for cluster 4 compared to the rest, this seems to indicate that the quenched galaxies here are possibly undergoing more significant interactions. This further motivates additional exploration of the merger histories of these populations in a future paper.

Examining the mass weighted ages of the stellar and halo components reveals a similar pattern as in case B: the redder and the more satellite dominated, the younger the halo is with respect to the stellar component ($7.08_{-1.30}^{+1.80} \ Gyr$  vs $10.60_{-0.86}^{+0.92} \ Gyr$ for cluster 2, $8.48 _{-1.12}^{+0.97} \ Gyr$ vs $9.33_{-1.09}^{+1.22} \ Gyr$ for cluster 4, $8.93 _{-0.57}^{+0.57} \ Gyr$ vs $7.22_{-0.78}^{+0.76} \ Gyr$ for cluster 3). Interestingly, even though cluster 2 is much more quenched, since it is largely satellite dominated, its mass weighted age for the halo is lower than cluster 4, which is more closely dominated by centrals. In addition, the stars in cluster 2 are older than cluster 4. This further motivates the need of a detailed study of the merger histories of these galaxies. However, we have already shown how this clustering is an effective proxy for extracting populations with very distinct features, both for the stellar and halo component.

Regarding the relationship between case E's clusters and case B's clusters, we examined the overlap fraction of these populations: what percentage of galaxies in a particular cluster for case E came from a particular cluster in case B. We found that $99\%>$ of galaxies in cluster 1 in case E came from cluster 2 in case B, in cluster 2 of case E $99\%$ of galaxies came from cluster 1 in case B, and an additional 1\% from cluster 3. Cluster 3 sourced 63\% from cluster 2, 37\% from cluster 3. For cluster 4, it sourced 49\% from cluster 1, 51\% from cluster 3 . This highlights that the galaxies identified in cluster 1, the very quenched population were a mostly satellite subset identified in case B's clustering.

It is worth noting that for future works, the implementation of normalization is particularly important. We normalized the smoothed star formation histories with a max-abs normalization as \textit{sklearn.prepreprocessing.normalize}, allows you to choose which axis to normalize along, while its Scaler object implementation does not for the Maximum Absolute Value scaling. Normalizing each star formation history between 0 and 1 produced the results here. With the Scaler object, it does not. This aligns with earlier caveats that clustering can produce unstable populations, so it is worth noting that the extremely quenched population is liable to change under the implementation of a specific normalization used.

\subsection{Links with known populations }

The largest takeaway is that the clustering is effective at identifying and navigating subpopulations of the red sequence, and isolating populations with a relatively unique set of formation histories. What is notable is the split in the red sequence of quiescent galaxies seems to be largely driven by the central/satellite status, and their ex-situ merger mass fraction, as demonstrated in case B, which explored populations produced from clustering NMF components of the star formation and metallicity histories of IllustrisTNG galaxies, into three selected populations. These populations are defined by a split between star forming and quiescent galaxies (cluster 2 vs clusters 1 and 3), and an additional split between the satellite dominated quiescent population of cluster 1 and the upper mass branch of the red sequence, cluster 3, characterised by its large ex-situ merger mass fraction (Figures \ref{fig:SFHZH-B},\ref{fig:fMBplot}). Cluster 1 is a population whose largest event for its dark matter halos occurs around 5-6 Gyr ago, with significant growth in the halo mass seen in the satellite galaxies of this satellite dominated population, largely after the population has begun to quench. By contrast, cluster 3, the more centrally dominated quiescent population, is characterized by a more gradual reduction in its star formation history, potentially indicative of the role major mergers might play in shaping the rate at which these galaxies might quench, as described by \citet{Hani_2020} and \citet{Quai_2021}.

Expanding the number of clusters for the same parameter space as above to four (case E), we find a parallel to the work of  \citet{Hani_2020} and \citet{Quai_2021} regarding the role of star formation histories and mergers, where for star forming galaxies, major mergers tend to have a significant impact in increasing the star formation rate in the period following a merger, albeit with the possibility of further hastening any quenching that might follow compared to their control group. This higher merger mass fraction might place a constraint on the local environment of these galaxies, as \citet{Oh_2019} finds that for clusters of galaxies, where only recently accreted galaxies show observational evidence of merger induced changes. For the former, satellite dominated quiescent population, similarities can be seen with \citet{donnari2020quenched} and \citet{Oh_2019}, where pre-processing and infall can help account for the quenching of these galaxies.

Little differentiation emerges within the blue cloud population, and instead we get three populations at three different stages of quenching. Comparing the mean star formation histories, we note a significant split across clusters 2,3,4.  While cluster 3 could be classified as a population that is not fully quenched, the quenching timescale between cluster 2 and cluster 4 can be clearly seen in Figure \ref{fig:SFHZH-E}. Cluster 2’s mean star formation seems to quench from 6-9 Gyr ago, while cluster 4’s starts quench around 2-5 Gyr ago. This older population is significantly dominated by satellites (87.95\% for cluster 2, vs 55.27\% for cluster 4,  37.24\% for cluster 3), indicating that cluster 2 is a population that is truly red-and-dead, and primarily satellite. Of note is how the key discriminants of a given galaxy in \citet{donnari2020quenched}'s work produce populations that align with ours: whether it quenched before infalling as a satellite, or following an infall event whose 16/84th percentile intervals for the star formation histories of cases B and E resemble periods when different populations of satellites quenching at different times in \citeauthor{donnari2020quenched}, see Figures \ref{fig:SFHZH-B} and \ref{fig:SFHZH-E}.

In both cases E and F, the addition of a fourth cluster reveals an extremely quenched, satellite dominated population inhabiting the red sequence. By contrast, doing so for the colour-based clusters, i.e. case D, reveals an interesting subset of the “blue cloud” population of star forming galaxies.

In addition to a late peak in the star formation and metallicity histories in Figure \ref{fig:SFHZH-D}, a few more features of this population stand out to reinforce this picture: we find that satellites in cluster 1 have a significantly larger ex-situ mass fraction from flybys (median: -2.45 for satellites, -3.08 for centrals, vs. medians of -3.09 for centrals and -3.27  for satellites in cluster 3) than the blue cluster, and that unlike any other population, the ex-situ fraction is larger for the satellite subpopulation than centrals. These ex-situ fractions are remarkably similar to the green valley population of cluster 4 (-2.58,-3.09 central and satellite medians).

Looking at the halo and stellar mass weighted ages, we find that cluster 1 has a mean halo mass weighted age of $8.34^{+1.25}_{-0.44} \ Gyr$ and $6.07 Gyr^{+1.19}_{-1.06} \ Gyr$ for the mean stellar mass weighted age, very similar to cluster 3’s  $9.17^{+0.55}_{-0.46} \ Gyr$ and $6.36^{+0.86}_{-0.84} \ Gyr$  respectively. Altogether, this paints the picture of this population being an interesting approximation of a post starburst galaxy: a galaxy where its star formation has just ceased, and is about to transition onto the red sequence, potentially with the ex-situ mass fraction information being a key indicator of the role of mergers in this transition. Given that this population was found using photometric colours, it demonstrates the potential of this method even in surveys where we would have no access to star formation and metallicity histories see \cite{Wild_2014}.

Clusters such as cluster 3 in case B and cluster 3 in case E, with galaxies with that inhabit the red sequence but have a larger ex-situ merger fraction than others, have been seemingly able to extend their star forming epoch as a consequence of these mergers, either through shocking existing gas, or through possible pre-processing and infall of neighbouring galaxies \citet{donnari2020quenched}. In either case, this link between mergers and delayed quenching is of interest, and additional works in the literature support the link between major mergers and their role in driving and suppressing star formation in these galaxies. Since this work is not primarily focused on merger histories, and rather just identifying distinct evolutionary histories, we present this topic as an area of interest for a follow up investigation.

\subsection{Areas of future work}

The absence of a “green valley” population is notable, as the clustering largely divides galaxies that might be considered green valley (i.e. the ones seen in case A) between clusters 1 and 3. The slow quenching of cluster 3 bears a notable similarity to the quenching described by \citet{Schawinski_2014} of quiescent galaxies. Moreover, the split evolutionary tracks identified in \citet{Trayford_2016} draw an interesting parallel as well to the variety of quenched populations found in cases B and E. Exploring the way these populations manifest in non-optical wavelengths would also be an area of interest, building upon the criticisms of the green valley presented by \citet{Eales_2018}.

Pertinent questions are raised in works like \citet{Weigel_2017}, \citet{10.1093/mnras/staa2607}, \citet{pawlik2019diverse}, and \citet{Wilkinson_2017} among others, such as the relevance of major mergers in driving the evolution of these galaxies, and the role AGNs might play in regulating the dispersal of gas in these galaxies. Of particular interest is \citet{donnari2020quenched}'s series of papers, which explored the role of pre-processing, and would serve as an effective and direct test of the ability of these sub-populations in replicating the results of simulation-based work using only "observable" parameters. In particular, points (i)-(iii) of \citet{donnari2020quenched}'s conclusion bear a striking resemblance to the sub-populations of galaxies we have described in cases B and E: that of lower mass, satellite dominated populations that are largely quenched, with an associated accretion time of 4-6 Gyr, possibly linked to an infall event, with the exception of bluer galaxies. They also identified galaxies along the upper mass range of the red sequence, where above $log{M_{*}/M_{\odot}} = 10$, the satellite population is largely quenched, further highlighting the similarities of those satellites and the satellite dominated sub-population from case B. Finally, to quote \citet{donnari2020quenched}: "While, as expected, the quenched fractions
of IllustrisTNG central and satellite galaxies are generally lower at
higher redshifts (Fig. 2), frequent manifestations of environmental
processes in hosts more massive than about $10^{13} M_{\odot}$ are already
in place at $z \sim 1$ and the bulk of the $z = 0$ group and cluster
quenched satellites ceased their star formation many billion years
ago: 6–10, 1–7, and 4–8 Gyr ago (16th–84th percentiles) for satellites
that quenched as centrals, in their current host, or as satellites before
falling into it, respectively." This describes populations that greatly resemble our own yet again, the population quenching at 1-7 Gyr ago resembles that of cluster 3 in case B: a central dominated host population, while the 6-10 and 4-8 Gyr ago intervals resemble cluster 1 in case B, or they can be further delineated when applying these intervals to case E. 

Their work's conclusion is in tandem with one of our own: that satellite galaxies have a diverse range of paths to quenching, and for us, in several different clusterings, that diversity is captured by some of our clusters. 

We have only briefly touched upon the merger histories of these clusters, and would encourage the reader to consider exploring some of the questions regarding the histories presented by these clusters. In particular, how closely do the identified sub-populations correlate with differences in environment? This is relevant given some identified links between environment and the number of associated mergers, especially for more massive galaxies \citep[e.g.][]{Yoon_2017}, which highlights how below masses of $log{M_{*}/M_{\odot}} \geq 11.0$ the number of mergers a given galaxy experiences ceases to correlate with local density. The application of these populations, when cross-referenced with environmental data from TNG-100, could help provide insights as to the extent these clusters might trace out the environmental dependence of these populations. 

Moreover, as we discussed in an earlier section, we have also traced out a testable shorthand for identifying potential post-starburst galaxies and generating broad predictions about their non-observable properties just from using photometric colours in our clustering. This population is satellite dominated, with a large ex-situ mass fraction. We will explore the link between traditional PSBs and this population in a followup paper.

\section{Conclusion}\label{sec:Conclusion}

We extracted a series of "observable" parameters from the IllustrisTNG-100 simulation, and clustered them with a Gaussian Mixture Model (GMM), studying a variety of parameterizations with different clusters. We focused on clustering with the photometric colours $u-g$,$g-r$, $r-i$ and NMF representations of the extracted star formation and metallicity histories. We ultimately found that in terms of effectively identifying populations with distinct evolutionary histories, the use of the star formation and metallicity histories was indispensable. Using these parameters in our clustering produced clusters with distinct histories associated with the dark matter halos of these galaxies. In particular, we found that cases B and E produced the most distinct of these evolutionary histories, as seen in our Discussion and Figures \ref{fig:SFHZH-B},\ref{fig:MAH_caseB},\ref{fig:fMBplot},\ref{fig:SFHZH-E}.

We highlight case B as the most significant sub-population of interest in this analysis, owing to distinct splits seen in the star formation histories and metallicity of the populations, this distinction extended to differences found in the halo and evolutionary components, i.e. the ex-situ mass fractions, the mass assembly history, and the mass weighted ages of the stellar and halo components. The most consequential split found in this three cluster case is between the populations that inhabit the red sequence: a lower mass, satellite dominated one, and a higher mass central dominated one. Taking what was discussed from the Discussion, we concluded that they had significantly different evolutionary histories.

Case E is interesting owing to the distinct populations it identifies primarily within the red sequence, and as an examplar of how adding an extra cluster to our algorithm alters the populations initially discussed in case B. We found three populations at varying degrees of quenching, and a single star forming population. Two of these populations, cluster 2 and cluster 4 are definitively quenched, and within distinct windows of time. The earlier it quenched, the more satellite dominated the population is. Interestingly, when comparing the mean mass weighted ages of the stellar and halo mass, these two populations have older stellar mass weighted ages than the halo mass weighted ages, while the opposite is true for the other two populations. This is indicative of a major difference in the role mergers play in the evolution of these populations.

Case D, while producing populations not as distinct as those in cases B and E, is notable in what we found in cluster 1, a population that could be potentially described as related to post-starburst galaxies (PSBs), given the spike in its star formation and metallicity histories, and we only needed photometric colours to identify it, and has some unique features in its ex-situ mass fraction and recent stellar evolutionary history.

We summarize the key conclusions below:

\begin{itemize}

\item The cases that made use of the star formation and metallicity histories in their clustering produced results that had  notable differences in the distribution of the in/ex-situ mass fraction, merger history and halo mass assembly history, indicative of the depth of the differences between clusters.

\item The populations identified in cases B,C,E,F also echo a point raised elsewhere in the literature, where satellite galaxies in particular seem to follow a diverse range of pathways to quenching in the present day, and that these pathways are reflected by the subpopulations identified in case E, and cases B,C to a lesser extent: as the 84/16th percentile intervals given by \citet{donnari2020quenched} closely resemble the ones seen for the quiescent populations of case E. This also suggests at the link between infall and the populations identified by these clusters.

\item The populations identified by our clustering show that the red sequence of quiescent galaxies is split into a central dominated subpopulation, and a satellite dominated one. That satellite dominated one can be further split if we examine these populations with four clusters instead of three (i.e. case E), with differences between the mass weighted ages of the halo and stellar component hinting at different modes of quenching.

\item While colour based clusterings did not produce populations that were as distinct as those in cases B,C,E,F, there are still populations of interest. Namely, case D, cluster 1, where clustering with broadband photometric colours for four populations produces a subset of star forming galaxies that are of potential observational interest. This cluster, cluster 1 is a subset of the star forming "blue cloud" population, but with relatively recent spike in its star formation and metallicity history (Figure \ref{fig:SFHZH-D}), and with considerable differences in its ex-situ mass fraction that could suggest at a merger/flyby rich subset of galaxies in the blue cloud. We argue that this population bears some notable similarities to PSB galaxies that have been found elsewhere in the literature, their precise relationship with PSBs needs to be characterized.

\item Our clusterings in cases B and E underscore that the most important and distinct divisions of galaxies do not necessarily reproduce a green valley population as identified elsewhere in the literature, rather it shows that the red sequence is an incredibly diverse population with a variety of evolutionary histories giving rise to those galaxies.

\end{itemize}


We have identified a number of populations that are of interest for future work, both in the context of other cosmological hydrodynamic simulations, where the populations we identified can be compared to the carefully selected ones seen in the literature \citep[e.g.][]{Hani_2020,Quai_2021,Wilkinson_2017}, or they can extended into an observational context, testing the general predictive power of these populations, and the simulation they were calibrated on, and juxtaposed with existing work that has attempted to do so \citet{donnari2020quenched}.

This work has demonstrated the the incorporation of quantities such as the star formation and metallicity histories of galaxies provides valuable insight into their evolutionary histories that exceeds what could be found using photometric colours. While these clusterings are not necessarily the best possible divisions of galaxies by evolutionary histories, they do have some distinct features that are of interest. The utility provided by these star formation histories and metallicity histories also underscores that the effort to extract those histories is not wasted, and that with these full histories of star formation and chemical enrichment, they can be leveraged to explore the evolutionary histories of these galaxies. Of course, it remains to be seen how an analysis on real data compares to the clusters we find here. We leave a detailed comparison between data and different cosmological simulations for a future paper.

\section*{Acknowledgements}

The authors would like to thank the referee for a thoughtful report that led to visible improvements in this paper,and also Dr. Jonas Chaves-Montero for their helpful feedback on star formation histories and their influence on photometric colours. We also want to thank Dr. Vivienne Wild for insightful commentary on PSB galaxies. 

\section*{Data Availability}
The data underlying this article is available in from  IllustrisTNG website \href{https://www.tng-project.org/data/docs/specifications/}{https://www.tng-project.org/data/docs/specifications/} data catalogues are freely available from their website, alongside their supplementary catalogues, which we made use of in this work. The mass assembly histories were computed by Harry Chittenden and have been submitted for publication \citet{Chittenden_2022}. \textit{sklearn} is a public python package, and was used for all the machine learning related tasks. A notebook with the data and code summarizing the figures from this work is available at the following repository: \href{https://github.com/tsfraser/FraserTojeiroChittendenPaper}{https://github.com/tsfraser/FraserTojeiroChittendenPaper}.




\bibliographystyle{mnras}
\bibliography{MSc_3}




\bsp	
\label{lastpage}
\end{document}